\documentclass[letterpaper,twocolumn,10pt]{article}
\usepackage{template/usenix/usenix}
\usepackage{booktabs}   
\usepackage{tikz}
\usepackage{filecontents}
\usepackage{amsmath,amsfonts,mathrsfs,amssymb}
\usepackage{graphicx,subfig}
\usepackage[normalem]{ulem}
\usepackage[linesnumbered,lined,boxed, commentsnumbered, ruled]{algorithm2e}%boxruled boxed
\usepackage{algorithmic}
\usepackage{multirow}
\usepackage{amsopn}
\usepackage{etoolbox}
\usepackage{xcolor}
\usepackage{url}

\usepackage{hyperref}
\usepackage{xspace}
\usepackage{indentfirst}
\usepackage{breakurl}
\usepackage{microtype}
\usepackage{color}
\usepackage{listings}
\usepackage{enumitem}
\usepackage{diagbox}
\usepackage{colortbl}
\usepackage{pifont}
\usepackage[misc]{ifsym}
\usepackage{ulem}
\makeatletter
\newcommand\figcaption{\def\@captype{figure}\caption}
\newcommand\tabcaption{\def\@captype{table}\caption}
\makeatother

\definecolor{amethyst}{rgb}{0.6, 0.4, 0.8}
\definecolor{byzantine}{rgb}{0.74, 0.2, 0.64}
\definecolor{cadmiumred}{rgb}{0.89, 0.0, 0.13}
\definecolor{brinkpink}{rgb}{1.0, 0.65, 0.79}%{0.98, 0.38, 0.5}

\definecolor{carnationpink}{rgb}{1.0, 0.65, 0.79}
\definecolor{caribbeangreen}{rgb}{0.67, 0.88, 0.69}
\definecolor{carrotorange}{rgb}{1.0, 0.66, 0.07}

\definecolor{alizarin}{rgb}{0.82, 0.1, 0.26}
\definecolor{ao(english)}{rgb}{0.0, 0.5, 0.0}
\definecolor{cadmiumorange}{rgb}{0.8, 0.33, 0.0}

\newcommand{\name}{{Minder}\xspace}
\newcommand{\etc}{\emph{etc.}\xspace}

\newcommand{\eg}{\emph{e.g.,}\xspace}
\newcommand{\etal}{\emph{et al.}\xspace}

\newcommand{\yangtao}[1]{\textcolor{black}{#1}}
\newcommand{\new}[1]{\textcolor{black}{#1}}

\if 0
\AtBeginDocument{%
  \providecommand\BibTeX{{%
    \normalfont B\kern-0.5em{\scshape i\kern-0.25em b}\kern-0.8em\TeX}}}
\fi

\begin{document}

\title{\name: Faulty Machine Detection for Large-scale Distributed Model Training}

\author
{
	{\rm Yangtao Deng$^1$, \hspace{0.2cm} Xiang Shi$^2$, \hspace{0.2cm} Zhuo Jiang$^2$, \hspace{0.2cm} Xingjian Zhang$^1$, \hspace{0.2cm} Lei Zhang$^2$} \\
  {\rm Zhang Zhang$^2$, \hspace{0.2cm} Bo Li$^2$, \hspace{0.2cm} Zuquan Song$^2$, \hspace{0.2cm} Hang Zhu$^2$, \hspace{0.2cm} Gaohong Liu$^2$} \\
  {\rm Fuliang Li$^3$, \hspace{0.2cm} Shuguang Wang$^2$, \hspace{0.2cm} Haibin Lin$^2$, \hspace{0.2cm} Jianxi Ye$^2$, \hspace{0.2cm} Minlan Yu$^4$} \\
	$^1$Tsinghua University \hspace{0.2cm}  $^2$ByteDance \hspace{0.2cm}  $^3$Northeastern University \hspace{0.2cm}  $^4$Harvard University
}

\pagenumbering{arabic}
% \pagenumbering{gobble}

\maketitle

\renewcommand{\thefootnote}{}
\footnotetext{The first two authors contributed equally to this paper. Zhuo Jiang and Minlan Yu are the corresponding authors. This work was done while Yangtao Deng and Xingjian Zhang were doing a joint research project at ByteDance.}

\renewcommand{\thefootnote}{\arabic{footnote}}

\begin{abstract}

\noindent
Large-scale distributed model training requires simultaneous training on up to thousands of machines. Faulty machine detection is critical when an unexpected fault occurs in a machine. From our experience, a training task can encounter two faults per day on average, possibly leading to a halt for hours. To address the drawbacks of the time-consuming and labor-intensive manual scrutiny, we propose \name, an automatic faulty machine detector for distributed training tasks. The key idea of \name is to automatically and efficiently detect faulty distinctive monitoring metric patterns, which could last for a period before the entire training task comes to a halt. \name has been deployed in our production environment for over one year, monitoring daily distributed training tasks where each involves up to thousands of machines. In our real-world fault detection scenarios, \name can accurately and efficiently react to faults within 3.6 seconds on average, with a precision of 0.904 and F1-score of 0.893.
\end{abstract}

\vspace{-0.1in}
\section{Introduction}
\label{sec:introduction}

\noindent
\yangtao{
Recent years have witnessed a rapid increase in dataset sizes and the number of parameters in models, especially in Large Language Models (LLMs). The GPT-4 model \cite{achiam2023gpt}, an instance of the Mixture-of-Experts (MoE) paradigm, demonstrates this growth with its 1.8T parameters. 
Other latest models also exhibit this trend, with parameter counts exceeding 500 billion \cite{smith2022using, chowdhery2022palm}. The feasibility of training such extensive models efficiently has been realized through large-scale machines and GPUs  \cite{295545,jiang2024megascale}. It has also been accompanied by advancements in distributed model training \cite{li2020pytorch,shoeybi2019megatron}, high-performance collective communication\cite{li2014communication, patarasuk2009bandwidth}, and fault-tolerant techniques \cite{he2023unicron}. A system of such vast size and complexity involves a huge amount of computation, communication, and storage resources as well as software support for a task. Consequently, the potential for faults is high, leading to the possibility of task failure.
}

\textit{Faulty machine detection} thus has become a significant bottleneck in the maintenance of distributed tasks. In our production environment, an accidental hardware or software fault occurs twice a day on average. The entire task may be forced to stop for hours or days until fixed for retraining. The economic loss for a customer can reach more than \$1700 in a 128-machine task for 40 minutes (case in \ref{subsec:anomalies}). Training a GPT-2 model with 1.5 billion parameters and 40GB dataset \cite{radford2019language}, for instance, takes 200 days utilizing an NVIDIA V100 GPU \cite{Nvidia_2H} (or 12 days for a DGX-2H server). If the training process is frequently interrupted by such faults, operating expenses and time costs will increase significantly.

However, the current manual diagnosis method is unsatisfactory. Once a halted task notification is received, the engineer needs to check the training parameters. Meanwhile, engineers from the training, physical networking, storage, and hardware teams, are also involved in diagnosis, since a fault can occur in any machine component. Examining machine logs and conducting offline performance tests on relevant hardware devices are required until the fault is detected (usually for hours). Delayed notifications, incomplete log content, and the complicated process of manual diagnosis amplify the unpredictability of time and labor costs.

It's necessary to design effective and accurate faulty machine detection methods that can quickly react to faults at runtime, not only providing better reliability but also eliminating manual efforts.
Achieving such goals is challenging because a machine can fail due to various types of faults. These faults can occur in any possible component, including hardware and software, and can be intra-host or inter-host. Besides, the abnormal pattern of monitoring metrics varies from task to task, making the traditional supervised anomaly detection methods impractical, because even the same behavior might be abnormal in a task with a different workload and machine scale. 
Additionally, there isn’t an individual monitoring metric that necessarily signals a fault. For instance, CPU or GPU usage is the most sensitive metric for fault indication, based on our observation from real production data. However, neither one is guaranteed to identify the faulty machine for Error Correction Code (ECC) errors. If noises exist in monitoring data, the detection may be even misguided. Therefore, faulty machine detection for distributed training is challenging. 

Instead of creating a monolithic predictor with available monitoring data, we developed \name by leveraging the ideas of similarity (\ref{subsec:similarity}), continuity (\ref{subsec:continuity}), training individual models for each monitoring metric (\ref{subsec:one_model}), and metric prioritization (\ref{subsec:metric_prioritization_sequence}). \name resolves the challenges by recognizing that a machine with a fault displays an abnormal pattern in certain metric data that differs from other machines and lasts for a duration. We also train individual models for data
denoising. We then track the dissimilarity between the denoised data for each machine and monitor its duration. By repeating this to individual metrics, the faulty machine is detected. To further expedite detection, we prioritize metrics to identify the most sensitive ones when a fault occurs. 

We designed, implemented, and deployed \name for all the distributed training tasks. \name operates without interrupting the running of the training machines, only requiring the pulling of monitoring data from the Data APIs for backend run-time analysis. Host metrics used by \name cover the aspects of computation, communication, and storage. Manual labors are released from the debugging process since \name can react within 3.6 seconds (\ref{subsec:overall_detection_performance}), reducing over 99\% of the time of manual debugging (shorter time by 500 ×). \name has an overall precision and F1-score of 0.904 and 0.893. 

We make the following contributions.
\begin{itemize}
\item \yangtao{An investigation of the fault types and their correlations with various monitoring metrics (\ref{subsec:observations}). We empirically explain why some metrics are more sensitive to faults and outline the challenges for the detection (\ref{subsec:challenges}).}
\item The ideas of similarity, continuity, denoising models, and metric prioritization for the design of \name (\ref{sec:overview}, \ref{sec:design}). Our thorough evaluation of \name's implementation (\ref{sec:implementation}) and and ablation experiments highlight its fast reaction, high accuracy, and proper design choices (\ref{sec:evaluation}). 
\item Lessons we encountered when deploying \name in practical (\ref{sec:discussion}). We also point out future directions.
\end{itemize}	
\vspace{-0.15in}
\section{Motivation}
%\section{Background and Related Work}
\label{sec:background_motivation}
\vspace{-0.05in}

\subsection{Negative Impacts of Faults in Real-World Production Environments}
\label{subsec:anomalies}
\noindent
Once an unexpected hardware or software fault occurs in a machine, it does harm to the entire distributed training task at the machine level.

\begin{figure}[tb]
	\centering
	\includegraphics[width=0.98\columnwidth]{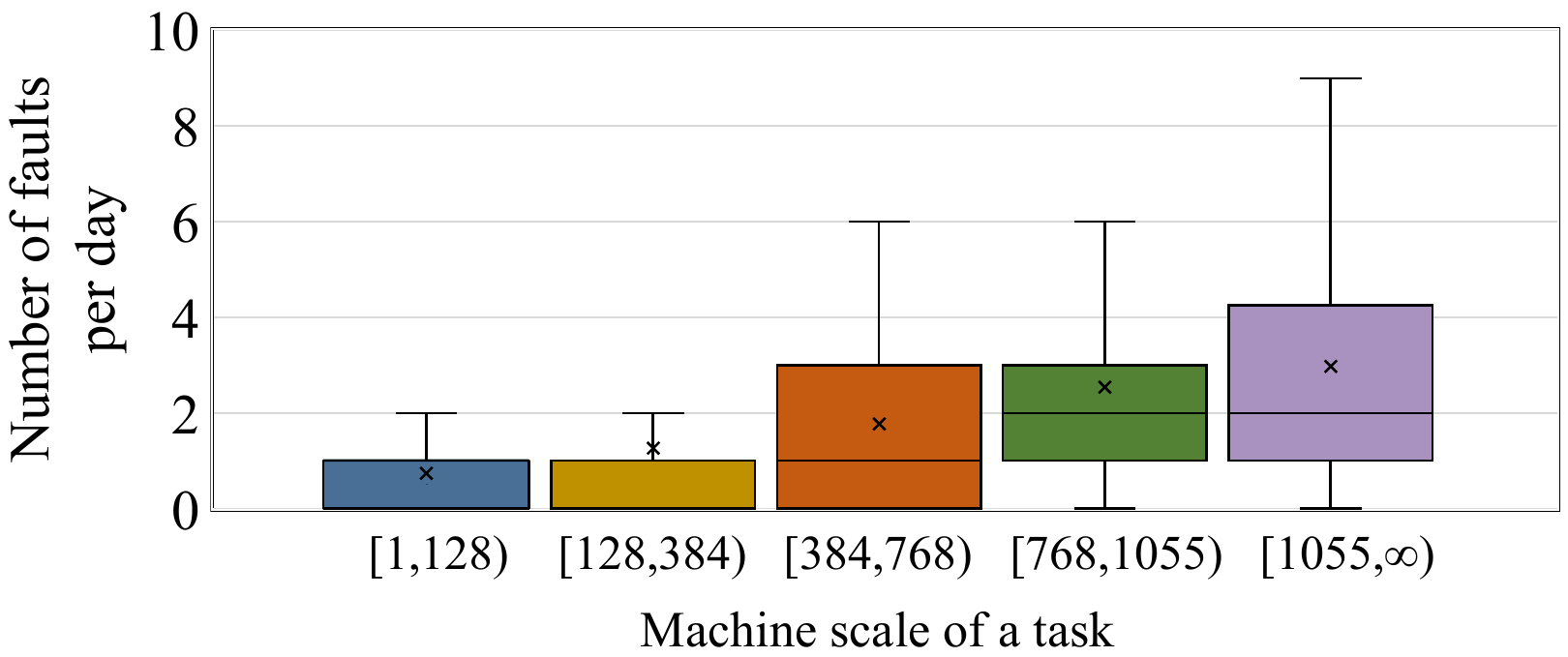}
	\vspace{-0.1in}
	\caption[12]{\yangtao{Fault frequency of tasks with different machine scale sizes.}}
	\vspace{-0.1in}
	\label{fig:motivation-machine_scale_failure_times}
\end{figure}

\noindent
\yangtao{\textbf{Faults are frequent due to the long training duration and large scale.} The relationship between a task's machine scale and the daily number of faults is illustrated in Figure~\ref{fig:motivation-machine_scale_failure_times}. Based on our seven-month statistics, the machine scale could significantly exceed 1024, training on more than 10000 GPUs. The occurrence of unexpected faults is highly correlated with the task scale, with an average of two faults a day. If we cannot react quickly to these incidents, such a frequent fault is a burden for training large models.}

\begin{figure}[t]
	\centering
	\includegraphics[width=0.95\columnwidth]{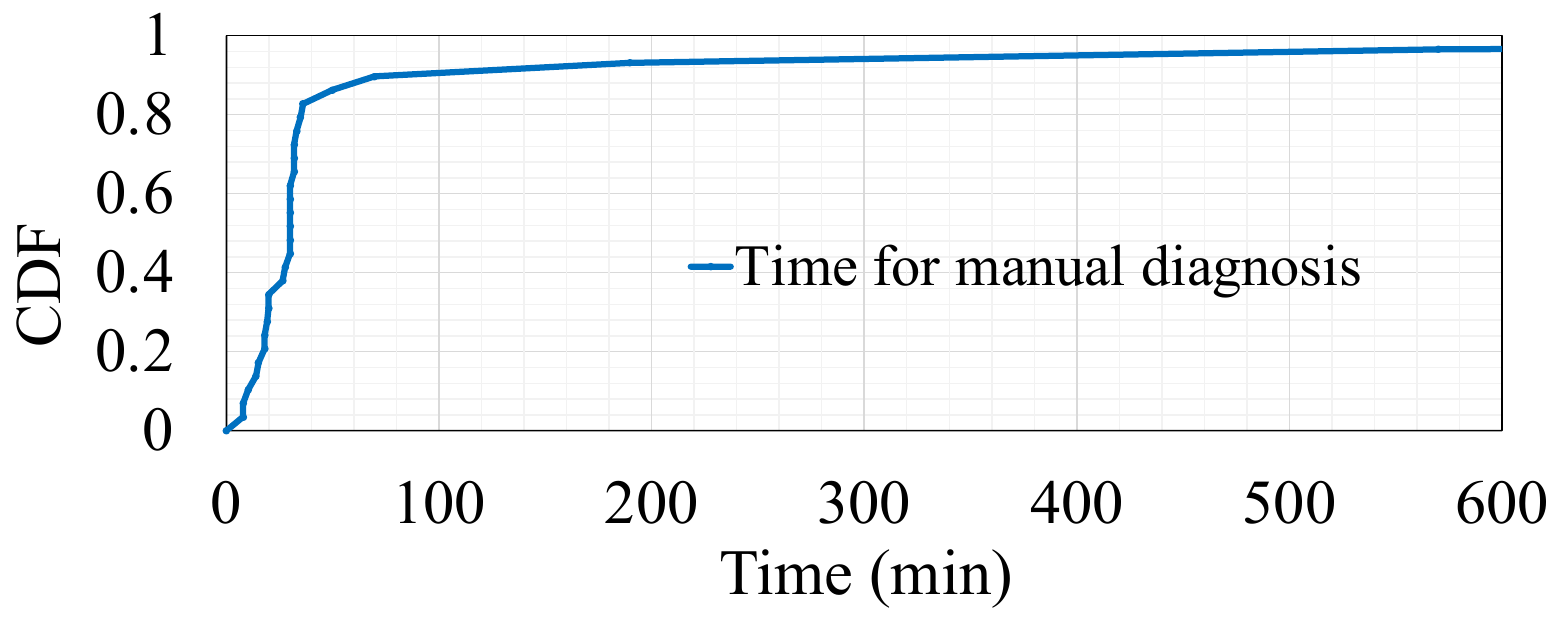}
	\vspace{-0.1in}
	\caption[12]{Time for task diagnosis in seven months.}
	% \vspace{-0.25in}
	\label{fig:motivation-manual_time}
         \vspace{-0.2in}
\end{figure}

\noindent
\textbf{A single fault can cause a large-scale task halt.} 
Based on empirical observations, a fault during the training process often originates as a host problem (\eg CUDA error, NVLink error \cite{GPU_metrics}). However, this issue can eventually cause a cascade effect, leading to the entire task being interrupted or a noticeable slowdown. For example, a hardware ECC error during communication will make the distributed systems interrupt (\eg when running in PS \cite{li2014communication} or All-reduce \cite{patarasuk2009bandwidth}) due to NCCL timeout \cite{Nvidia_NCCL_timeout} or network disconnections, thereby could resulting in more than half thousand of machines to stop and be in idle state in the production environment.

\noindent
\textbf{Faults take a long time to diagnose, increasing labor and resource costs.} Based on seven-month data in 2023, the time until the faulty machine can be manually diagnosed is shown in Figure~\ref{fig:motivation-manual_time}. The time lasts over half an hour on average and can be days. It ultimately results in GPU and NIC resources being left idle, leading to additional time costs upon resolution. More delays can result in significantly increased costs and time requirements for users.

\noindent
\textbf{An example: PCIe downgrading.} We highlight a real-world instance where a 128-machine training was forced to slow down severely for 40 minutes due to a PCIe degradation (PCIe is an intra-host, high-speed serial bus that connects a variety of devices, such as graphics cards, NICs, and solid-state drives). The faulty machine encountered PCIe degradation from 6.4Gbps to 4Gbps. The congested communication ultimately results in the underutilization of computational resources, where thousands of V100 GPUs were compelled to remain slowed down for 40 minutes. Such cost wastes for customers can be \$650 for 40 minutes (given the public renting price of \$2.48 per GPU for an NVIDIA V100 \cite{GPU_pricing}).

\subsection{Today's Solution and Drawbacks}
\label{subsec:today_solution}
\noindent
A straightforward approach after an unexpected fault occurs is inspecting the existing hardware and software logs on the machines, as well as \texttt{dmesg} \cite{dmesg}, \texttt{netstat}, \texttt{top} commands. However, this process involves the following three limitations.

\noindent
\textbf{The notification of when to trigger a diagnosis is not timely.} Firstly, engineers are only alerted once the task has stopped entirely. Certain faults will slow down the training speed in Model FLOPs Utilization (MFU) but are not severe enough to stop the task. The current approach fails to detect  degradation in performance as long as the task continues to run. The task will continue running with deteriorated performance for a period in the PCIe downgrading example. 

\noindent
\textbf{Scrutinized content is incomplete or redundant.} Upon a fault notification, the logs or counters recorded during training will be reviewed. The logs are typically maintained in plain text, including built-in software-layer logs (\eg NCCL and CUDA logs), hardware-layer logs, and network logs. 
Previous knowledge and experience guide the decision on which logs to include and check, resulting in some log content being overlooked. Meanwhile, logs do not include monitoring metrics like GPU power, temperature, and NVLink bandwidth. In the PCIe degradation case (\ref{subsec:anomalies}), identifying the faulty machine was hard as critical monitoring data, like Priority-based Flow Control (PFC) packet rates, was not promptly inspected from the logs. Besides, the log content frequently includes redundant data, such as environmental parameters and warnings. The detection time will be lengthened.

\noindent
\textbf{The diagnosis analysis is a complicated and time-consuming process.} 
Engineers from multiple teams are involved in the diagnosing process. Figure~\ref{fig:motivation-manual_time} presents the time-consuming diagnosis process. It could take as long as several days to detect the faulty machine.

\noindent
\textbf{Manual diagnosis procedure and fault propagation for the PCIe downgrading case (\ref{subsec:anomalies}).} The detection took 40 minutes in total and involved multiple teams. The engineer in charge scrutinized model-related information, parallelism settings \cite{jiang2024megascale}, dependencies, environmental, and framework (\eg Megatron-LM \cite{shoeybi2019megatron}) parameters. Meanwhile, the network team scrutinized intra-host throughput, Remote Direct Memory Access (RDMA) traffic, packet loss/randomness, congestion indicators, drivers, and routing. The storage and hardware teams inspected HDFS\&SSD usage, GPU\&CPU usage, NIC health, and machine scheduling.

The fault propagation initiated from PCIe downgrading to PFC surge. The NIC buffer of the faulty machine was filled after the PCIe degraded. The consequent bottleneck inter-host communication caused a PFC Tx packet surge. The congestion also raised both Explicit Congestion Notifications (ECN) received and Congestion Notification Packets (CNP) sent \cite{ECN_CNP}. As a result, the NIC throughput across all machines dropped from 6.5Gbps to 4.9Gbps. Reduced computation data led to declined GPU tensor core usage \cite{gpu_tensor_core_utility}. Thus, the training performance was downgraded.

\definecolor{darkred}{RGB}{139,0,0}
% \textcolor{darkred}

\begin{table*}[htbp]
  \newcolumntype{P}[1]{>{\centering\arraybackslash}p{#1}}
  \small 
  \centering
  \caption{Fault types and the proportion of instances for each fault type being indicated by a metric.}
  \vspace{-0.1in}
  \label{failure_type}
  \begin{tabular}{|p{28mm}c|c|cccP{15mm}cc|}
    \hline
    \multicolumn{2}{|c|}{\multirow{2}{*}{\textbf{Fault type}}} & \multicolumn{1}{c|}{\multirow{2}{*}{\parbox{20mm}{\textbf{Frequency of each fault type}}}} & \multicolumn{6}{c|}{\textbf{Metrics}} \\
          &    &   & \textbf{\footnotesize CPU} & \textbf{\footnotesize GPU} & \textbf{\footnotesize PFC} & \textbf{\footnotesize Throughput} & \textbf{\footnotesize Disk} & \textbf{\footnotesize Memory}  \\
    \hline
    \multirow{7}{*}{\parbox{28mm}{\textbf{Intra-host hardware faults (55.8\%)}}}
     & \textbf{ECC error} & \textbf{38.9\%} & \textcolor{darkred}{80.0\%} & 65.7\% & 8.6\% & 45.7\% & 11.4\% & 57.1\% \\ 
        & \textbf{PCIe downgrading} & \textbf{6.6\%}   & 0.0\% & 8.3\% & \textcolor{darkred}{100\%} & 33.3\% & 8.3\% & 0.0\% \\ 
        & \textbf{NIC dropout} & \textbf{5.7\%} & \textcolor{darkred}{100\%} & \textcolor{darkred}{100\%} & 0.0\% & \textcolor{darkred}{100\%} & 0.0\% & \textcolor{darkred}{100\%} \\ 
        & \textbf{GPU card drop} & \textbf{2.0\%} & \textcolor{darkred}{75.0\%} & \textcolor{darkred}{70.0\%} & 5.0\% & 50.0\% & 20.0\% & 55.0\% \\ 
        & \textbf{NVLink error} & \textbf{1.7\%} & \textcolor{darkred}{83.3\%} & 50.0\% & 16.7\% & 50.0\% & 0.0\% & 66.7\% \\ 
        & \textbf{AOC error} & \textbf{0.9\%} & 25.0\% & 25.0\% & 0.0\% & 25.0\% & 25.0\% & 25.0\% \\
    \hline
    \multirow{2}{*}{\parbox{28mm}{\textbf{Intra-host software faults (28.0\%)}}} 
    & \textbf{CUDA execution error} & \textbf{14.6\%}  & 61.9\% & 57.1\% & 19.0\%  & 33.3\% & 14.3\% & 61.9\%      \\
        & \textbf{GPU execution error} & \textbf{7.7\%} & 50.0\% & \textcolor{darkred}{71.4\%} & 14.3\% & 42.9\% & 21.4\% & 42.8\% \\  
         & \textbf{HDFS error} & \textbf{5.7\%} & 57.1\% & 57.1\% & 0.0\%  & 14.3\% & 0\%  & 14.3\% \\  
    \hline
    \parbox[c]{28mm}{\vspace{0.5mm}\textbf{Inter-host network faults (6.0\%)}}
    & \textbf{Machine unreachable} & \textbf{6.0\%} & 47.4\% & 63.2\% & 0.0\% & 53.6\% & 26.3\% & 15.8\% \\ 
    \hline
    \textbf{Others (10.3\%)} & \large - & \textbf{10.3\%} & \large - & \large - & \large - & \large - & \large - & \large - \\ 
    \hline
    \end{tabular}
    \caption*{\footnotesize Note: the introduction of each fault type is in Appendix~\ref{sec:Appendix_introduction}.}
    \vspace{-0.3in}
\end{table*}

\subsection{Real-world Faulty Case Studies}
\label{subsec:observations}
\noindent 
\yangtao{To address the drawbacks of the manual log analysis, we first conducted an in-depth review of the fault types that occurred over seven months. 
Table~\ref{failure_type} shows the common types of faults, their frequencies, and the proportion of instances for each fault type that a metric could indicate. The proportions are determined empirically by examining the available instances and quantifying the number that exhibited abnormal patterns in the monitoring data following a fault. The monitoring data contains multiple metrics and is sampled per second.}

\yangtao{We come up with the following observations. Firstly, hardware faults make up the majority of faults (55.8\%), in which ECC errors constitute a large proportion (38.9\%). Errors that happened in CUDA or GPU also make up a large proportion. Unfortunately, these errors are hard to predict or avoid.}

\yangtao{Moreover, each metric displays varying probabilities of indicating a type of fault. There isn't a single metric that effectively signals all of them. ECC error, NIC dropout, GPU card drop, NVLink error, CUDA, and GPU execution error strongly correlate with CPU or GPU metrics. Likewise, PCIe downgrading, NIC dropout, and machine unreachable are most relevant to network metrics, such as PFC and throughput. We explain these observations from our experience. For the CPU metric, a fault on a machine will cease the CPU process, reducing its CPU usage. However, the other machines retain their processes for a while, waiting for data synchronization, due to the Kubernetes management, NCCL timeout setting, and heartbeat mechanism setting with a predefined time window. Thus, these machines maintain normal CPU usage before the task halts. For the GPU metric, a GPU drop or process kill during computation leads to low GPU usage. Conversely, other machines carry on running their CUDA kernels before the timeout and heartbeat check, maintaining normal GPU usage. For the PFC metric, PFC signals on the faulty machine surge abruptly when the NIC buffer is filled, due to congestion-related network errors. Other machines exhibit a low number of PFC packets. The same trend applies to memory usage. Nonetheless, disk usage does not exhibit significant fluctuations based on our experience. Therefore, most fault types strongly correlate with CPU, GPU, Memory, and PCIe metrics. As an exception, AOC errors happen on the switch or the machine-side cables. If a switch AOC error occurs, machines connected to this switch port will be affected instantly. Such a large scale of affected machines can quickly propagate adverse effects to others. It is hard to capture abnormal patterns with second-level monitoring data. }

\subsection{Challenges}
\label{subsec:challenges}
\noindent
Based on the observations, we discovered that the following challenges must be addressed:

\noindent
\textbf{Challenge 1: Any machine could fail in various ways.} Advanced machines, such as Nvidia DGX-A100 \cite{Nvidia_A100}, incorporate as many as 8 Nvidia A100 GPUs and 4 Mellanox 200 Gb/s RDMA NICs (RNIC), all of which are potential fault points. As presented in Table~\ref{failure_type}, a fault could happen from intra-host computation, and communication, to inter-host networks, or from hardware (GPU, CPU, PCIe, NVLink, RNIC, memory, and disk) to software communication libraries (\eg NCCL), and training frameworks (\eg CUDA). It is hard to detect a faulty machine under all the unknown circumstances.

\noindent
\textbf{Challenge 2: The normal state of a monitoring metric is task-dependent.} Training tasks have various machine scales, data sizes, models, and training frameworks. Thus, a monitoring metric may have different normal states for various tasks.
For instance, a GPU temperature of 70 degrees Celsius is abnormal where GPU clock frequency is 1350MHz, but is regarded as normal in a task where GPUs work with a high clock frequency of 1800MHz. Traditional supervised anomaly detection is inappropriate for differentiating between normal and faulty machine states because the same input monitoring data can be labeled as either normal or abnormal.

\noindent
\textbf{Challenge 3: The correlation between fault types and monitoring metrics is not necessarily one-to-one.} \yangtao{On one hand, a metric anomaly may be caused by various fault types. For example, a decrease in CPU usage could be caused by ECC error, NVLink error, and other faults. On the other hand, for a fault type, there isn’t a single metric that necessarily signals it. As an example in Table~\ref{failure_type}, ECC errors could potentially be noticed by either CPU or GPU usage of the faulty machine, yet neither metric guarantees nor necessarily both. Thus, a single fault type can be manifested via multiple abnormal metrics, with no solitary metric providing a guaranteed indication. Instead, the metrics exhibit an "or" correlation for a fault type. Consequently, we cannot simply use a model that incorporates the data from all metrics. Rather, we use individual per-metric models for detection (analysis in ~\ref{subsec:model_comparison}).}

\noindent
\textbf{Challenge 4: Noises exist in time series monitoring data.} The monitoring data inevitably consists of noises due to jitters, inaccurate sensors, temperature, timestamp misalignment, network interruptions, or other issues. Short-term noises will mislead us to regard a machine as the faulty one, resulting in extra time and labor burden. Thus, the raw data cannot be used directly for detection (analysis in \ref{subsec:model_comparison}).

\vspace{-0.1in}
\section{Design Overview}
\label{sec:overview}
\vspace{-0.09in}

\noindent
\yangtao{In this section, we introduce four design choices to address the challenges, revolving around \name: an automatic, responsive, and accurate watcher to detect the machine with an unpredicted fault during distributed training that leads to a task halt. We examine the dissimilarity (\ref{subsec:similarity}) among machines. Then we evaluate the continuity (\ref{subsec:continuity}) of the detected faulty machine candidate for filtering out bursty jitters. Since metric data contains noises, we use models to denoise and reconstruct the raw data. Specifically, individual models corresponding to various monitoring metrics are trained, instead of integrating them into a single model (\ref{subsec:one_model}). We also prioritize the metrics (\ref{subsec:metric_prioritization_sequence}). The faulty machine can be detected swiftly by the top-prioritized metrics and corresponding models.}

\begin{figure}[tb]
	\centering
	\includegraphics[width=0.95\columnwidth]{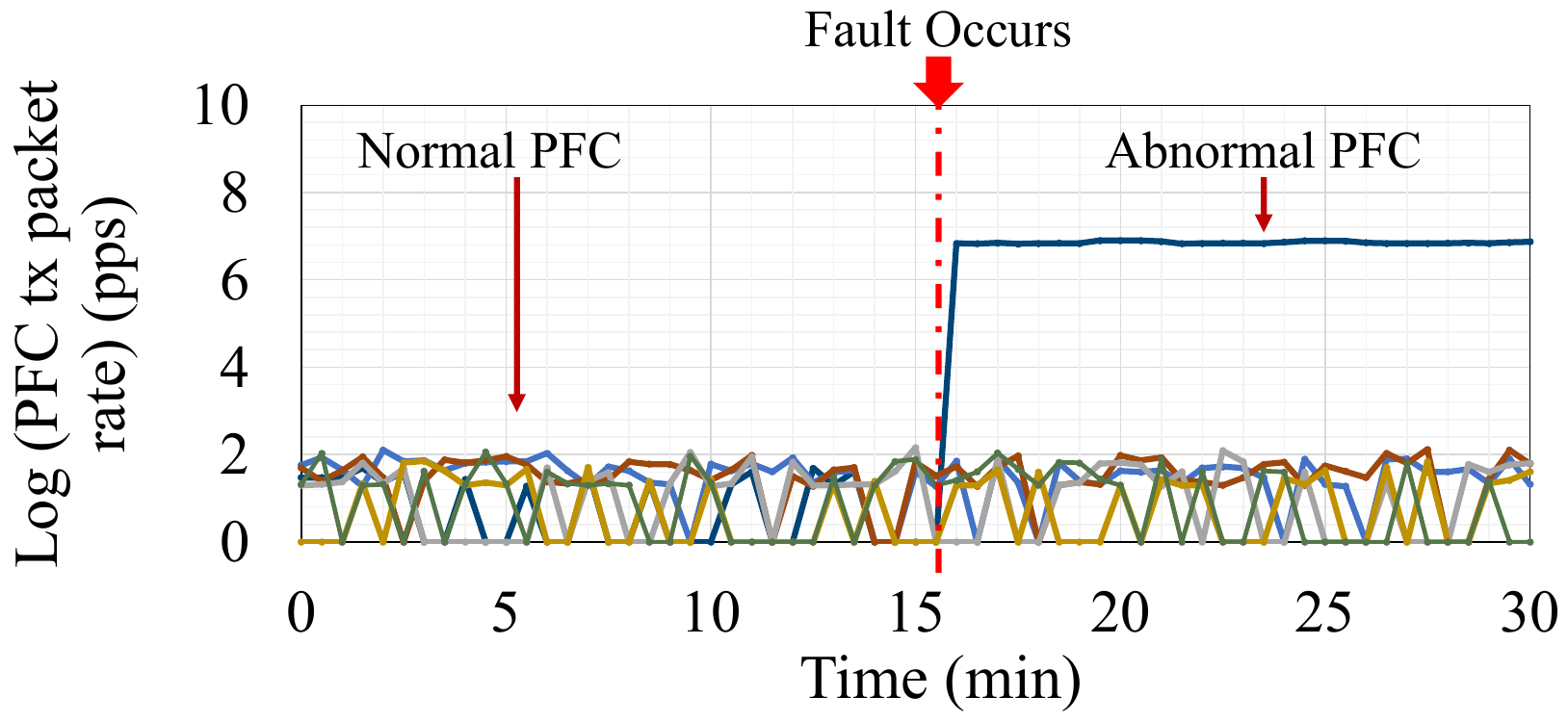}
	\vspace{-0.1in}
	\caption[12]{PFC tx packet rate pattern for each machine before and after a fault occurs.}
	\label{fig:motivation-pfc}
 	\vspace{-0.25in}
\end{figure}

\subsection{Machine-level Similarity}
\label{subsec:similarity}
\noindent
\yangtao{To address challenges 1 and 2, we notice that machine-level metric data exhibit similar behavior in parallel distributed training \cite{shoeybi2019megatron} at the second level.}

\yangtao{Data parallelism (DP) operates by splitting data equally among GPUs that store duplicate model parameters and optimizer states. Pipeline parallelism (PP) assigns model layers to multiple GPUs in a pipelined manner. In tensor parallelism (TP), each GPU executes a portion of a computation process to improve hardware utilization in parallel.}

\yangtao{By integrating these techniques into a 3D parallelism framework, large-scale model training, such as LLM training \cite{jiang2024megascale}  or multi-modal training, can be effectively facilitated. TP is typically constrained within a single machine, whereas DP or PP groups involve inter-host communications. The computation, storage, and communication loads are evenly balanced across machines at the second level. Therefore, similar monitoring data fluctuations will be observed across machines. In the PCIe downgrading example, the initial \texttt{PFC Tx Packet Rate} patterns are notably uniform for all the machines in Figure~\ref{fig:motivation-pfc}. However, if a machine undergoes a fault, its monitoring data will display distinctive differences, offering an opportunity for detection. This principle can be extended to other metrics, as demonstrated by the possibilities offered in Table~\ref{failure_type}.
Hence, the denoised metric data from each machine is used for calculating the similarity with others. The machine identified with the longest "distance" from others could be assumed to be faulty, regardless of the fault type.}

\textbf{Why not use a supervised learning model for faulty machine detection?} Unlike many supervised learning-based anomaly detection approaches~\cite{gao2020scouts, laptev2015generic, arzani2020privateeye, liu2015opprentice}, \name uses unsupervised learning and similarity-based distance check, due to the different problem contexts. Firstly, labeling the data as normal or abnormal is impractical. As mentioned in challenge 2, the abnormal pattern is task-dependent. Different tasks present different normal ranges for the same monitoring metric under varying working conditions. Secondly, our objective is to identify which machine is to be blamed for the unexpected fault. This is not merely a classification problem of distinguishing normal or abnormal cases. Thus, developing a universal model via supervised training is challenging.

\subsection{Machine-level Continuity}
\label{subsec:continuity}
\begin{figure}[tb]
	\centering
	\includegraphics[width=0.9\columnwidth]{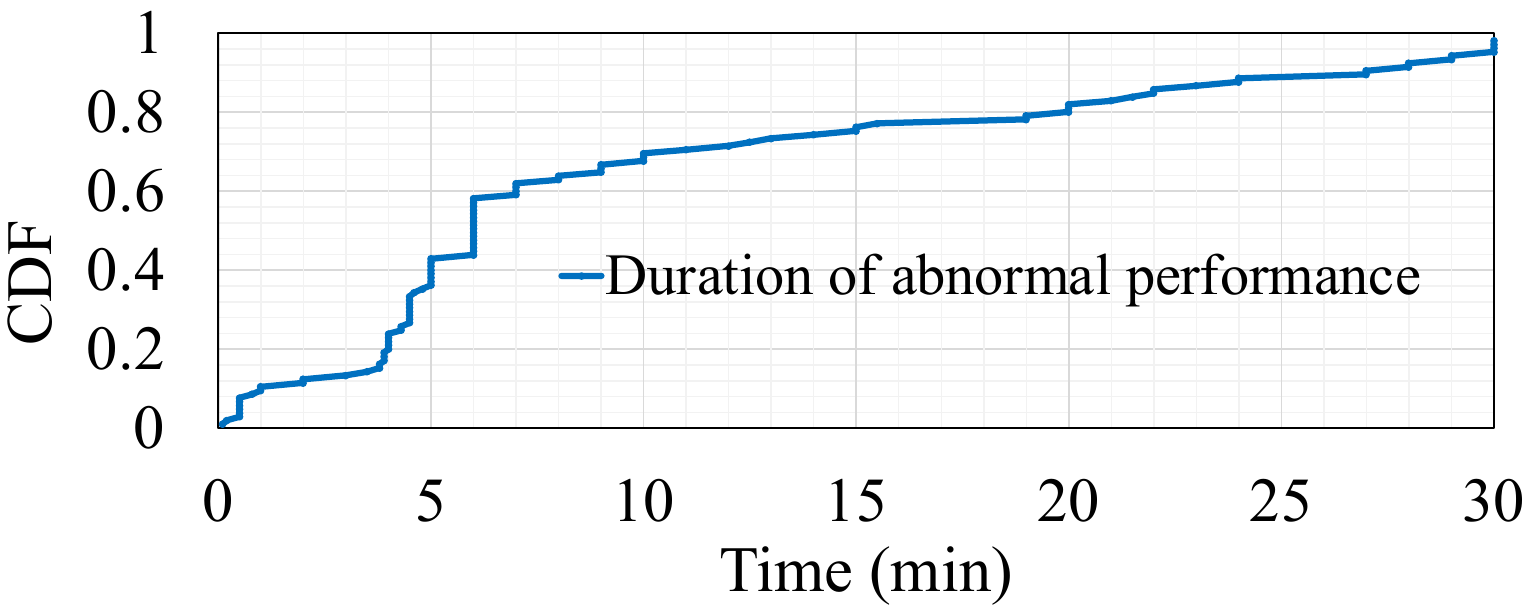}
	\vspace{-0.13in}
	\caption[12]{\yangtao{Duration of abnormal performance following a fault.}}
	% \vspace{-0.25in}
	\label{fig:motivation-duration}
         \vspace{-0.2in}
\end{figure}

\noindent
To further tackle challenge 2, we utilize the notion of abnormality continuity. Typically, abnormal performance persists for a few minutes upon a fault, but jitters typically last for a short duration. As illustrated in Figure~\ref{fig:motivation-pfc}, the machine with PCIe degradation experienced significantly higher \texttt{PFC Tx Packet Rate} for a period compared to other machines. \yangtao{By inspecting fault instances from seven-month data in 2023, the duration of abnormal performance after a fault occurs is depicted in Figure~\ref{fig:motivation-duration}. Most abnormal patterns last for over five minutes. Thus, if we recognize a machine displaying such dissimilarity continuously for a period, the machine may be faulty. In the case of raw data containing bursty noises, they will be filtered out (as analyzed in \ref{subsec:analysis_of_continuity}).}

\subsection{Individual Learning-Based Denoising Models for Each Monitoring Metric}
\label{subsec:one_model}
\noindent
\yangtao{For challenge 4, we utilize simple learning models for data denoising and reconstruction, in addition to basic data alignment and normalization. Variational autoencoder (VAE) and other generative probabilistic models are recognized for learning time-series data patterns and features~\cite{lin2020anomaly, su2019robust}. They are also known for learning embedding schemes that can infer the generation factors for most of the training data. This makes unsupervised learning particularly suited for modeling normal behavior in an anomaly detection task. As such, raw data is denoised and reconstructed by our learning-based models before being fed into further detection (as analyzed in \ref{subsec:model_comparison}).}

\yangtao{Meanwhile, based on challenge 3, no single metric provides a guaranteed indication for a specific fault type. The indication probability by a metric varies across different types of faults. For this reason, we opt for training individual models for each monitoring metric. These models and their corresponding metric data are used independently for denoising, similarity, and continuity detection.}

\yangtao{We do not merge all potential metrics into a single model for two main reasons. First, the time series of multiple metrics do not fluctuate in the same manner when a fault arises. Moreover, metrics' indication capacity differs even for a particular fault type, and one metric's capability varies across different types. As a result, integrating them into one model could lead to the model being misdirected or confused by the array of metrics (as analyzed in \ref{subsec:model_comparison}).}

\subsection{Prioritized Metric Sequence}
\label{subsec:metric_prioritization_sequence}
\noindent
\yangtao{To expedite detection, we prioritize metrics and only use the models of the top ones, since plenty of metrics (in Appendix~\ref{sec:Appendix}) could be collected and each could be trained with a model. This ensures we use the metrics with higher indication probabilities earlier. The faulty machine could therefore be detected sooner. If a model and its associated metric data cannot detect a faulty machine, we then move to the next metric and its model, following the prioritization result. We repeat this process until a faulty machine is identified.}
\vspace{-0.1in}
\section{\name Framework}
\label{sec:design}
\vspace{-0.09in}

\begin{figure}[tb]
    \vspace{0.00in}
	\centering
	\includegraphics[width=0.8\columnwidth]{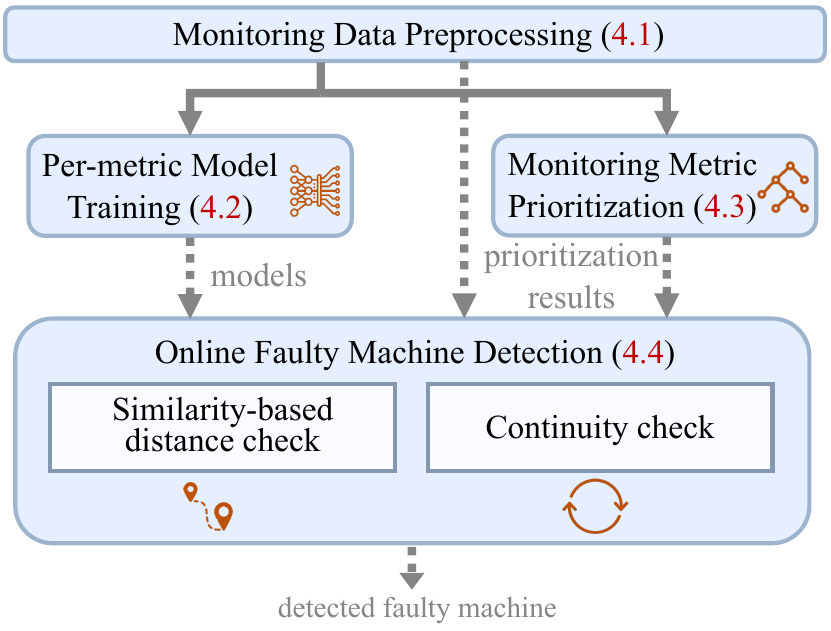}
	\vspace{-0.1in}
	\caption[12]{\yangtao{System architecture of \name.}}
	\vspace{-0.25in}
	\label{fig:design-framework}
\end{figure}

\noindent
\yangtao{The architecture of \name is shown in Figure~\ref{fig:design-framework}. Preprocessing (\ref{subsec:preprocessing}) is required for raw data from each machine. Per-metric Model Training (\ref{subsec:training}) and Monitoring Metric Prioritization (\ref{subsec:kpi_selection}) will train models and prioritize the metrics in their sensitivity to faults respectively, as two independent processes. The models and prioritization results are then used in Online Faulty Machine Detection (\ref{subsec:detection}) for run-time detection.}

\subsection{Preprocessing}
\label{subsec:preprocessing}
\noindent
Given a stream of monitoring data from each machine, \name needs to aggregate them into a series of time windows. Within each window, \name can do data denoising and machine-level similarity check on a set of metrics. By checking the anomaly continuity from consecutive time windows, \name detects the faulty machine.

Thus, \name preprocesses the collected monitoring data if it lacks alignment among certain metrics. \name first aligns the sampling points across all machines based on the corresponding sampling timestamps. If sample points are missed, \name uses data from the nearest sampling time for padding.

Normalization is adopted to ensure that the multi-dimensional monitoring data is integrated into an even distribution. \name normalizes the monitoring data based on the upper and lower limits of each metric, using the Min-Max normalization technique.

\subsection{Per-metric Model Training}
\label{subsec:training}
\noindent
\yangtao{As specified in \ref{subsec:one_model}, we train models for the learning-based denoising and reconstruction for subsequent detection. Since no single metric provides a guaranteed indication and a model incorporating various metrics might be misdirected, individual models should be trained for each metric.}

The preprocessed per-machine data within a time window is used as input instances to train an unsupervised model. For example, to train a model for \texttt{CPU Usage}, we use \texttt{CPU Usage} sample data within a time window with a length of $w$ (\eg 8) and a stride of 1 from each machine of the task. Multiple $1 \times w$ vectors are fed into the model respectively for training. Models for \texttt{CPU Usage}, \texttt{PFC Packet Rates}, and so on are individually trained. The parameters in the model include $hidden\_size$ (\eg 4), $latent\_size$ (\eg 8), and $lstm\_layer$ (\eg 1). With a model, the time series input of a monitoring metric from a machine could be reconstructed as a denoised vector for this machine.

\noindent
\textbf{Training choice: LSTM-VAE.} Specifically, VAE models are trained for \name.
VAE is widely used for denoising \cite{su2019robust} and compression of high-dimensional features \cite{sun2021ctf}, where normal vectors will be reconstructed into similar embeddings while abnormal ones will be reshaped into a more distinctive outlier. This highly effective unsupervised DL technique can reconstruct time series inputs into an arbitrary latent representation without sacrificing original characteristics. Consequently, VAE can enhance the accuracy and robustness of anomaly detection where labeling is absent \cite{xu2018unsupervised}. When the input consists mainly of normal training state vectors and only a small proportion corresponds to a faulty period, the VAE learns the vector distribution and denoises the jitters.

\begin{figure}
	\centering
	\includegraphics[width=0.95\columnwidth]{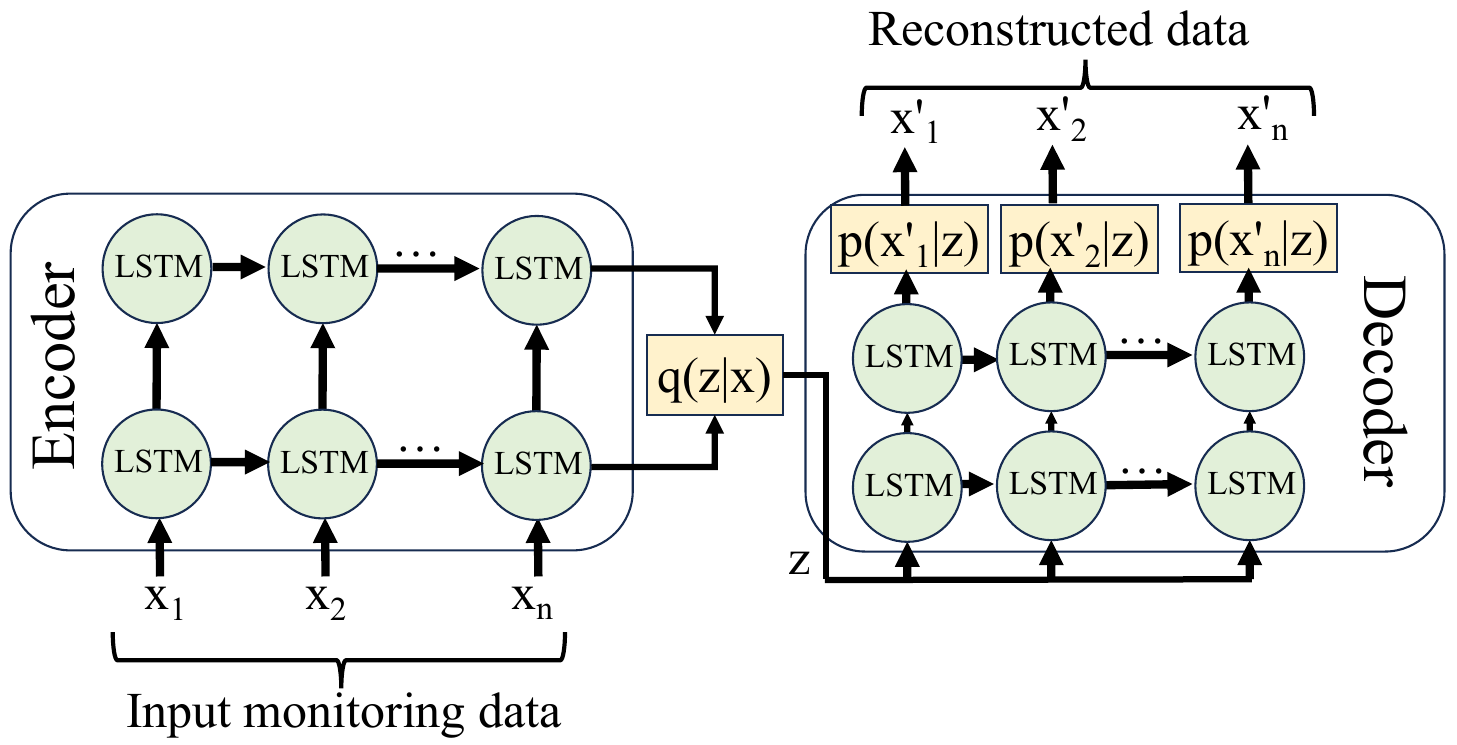}
	\vspace{-0.1in}
	\caption[12]{The LSTM-VAE structure for \name.}
	\vspace{-0.25in}
	\label{fig:design-vae}
\end{figure}

As depicted in Figure~\ref{fig:design-vae}, the VAE comprises an encoder and a decoder. The encoder extracts temporal features into a latent space embedding $z$. Subsequently, the decoder utilizes $z$ to restore the data to a new dimension output as a reconstruction of the distribution. Various statistical or machine learning techniques can be employed in the encoder and decoder to determine the optimal distributions. Given that our data is temporal time series, we utilize LSTM as both the encoder and decoder to extract temporal characteristics \cite{lin2020anomaly}. LSTM considers both forward and backward information of a time series to obtain complete correspondence information. As such, LSTM is an ideal choice for VAE.

\subsection{Monitoring Metric Prioritization}
\label{subsec:kpi_selection}
\noindent
\yangtao{As introduced in \ref{subsec:metric_prioritization_sequence}, we aim to use only the most representative metrics and their models for quick faulty machine detection. Given a large number of metrics, pinpointing the metrics more sensitive to faults is critical. By using the top prioritized metrics and their associated models first, the faulty machine will be detected more quickly.}

\yangtao{Consequently, following the steps below, \name will generate a prioritized list of metrics by their sensitivity to faults. The prioritization results can then be used in run-time detection, specifying which metrics and their models should be used first. Note that this process runs in parallel with the model training process in \ref{subsec:training}.}

\noindent
\textbf{Step 1: Z-score calculation for evaluating metric sensitivity to faults.} To identify the most sensitive metrics, \name utilizes the Z-score \cite{cheadle2003analysis}, because it depicts the dispersion of data distribution. A metric with a higher Z-score relates to an imbalanced distribution, where a faulty machine shows a dissimilar pattern from others. \yangtao{For a monitoring metric, the Z-score is computed for each machine at a sampling data point from the preprocessed data. For the $j$-th monitoring metric:}
$$Z_{ij}=\frac{x_{ij}-\bar{x_j}}{s_j}$$
where $Z_{ij}$ is the Z-score of the $i$-th machine, $x_{ij}$ is the sample value of the $i$-th machine, while $\bar{x_j}$ and $s_j$ are the average value and the standard deviation of all machines on the $j$-th metric. When a fault occurs, the affected machine exhibits abnormal behavior, leading to outlier samples with high Z-scores. 

For a time window of a training task, we use $max(Z_{ij})$ across all the machines for the $j$-th monitoring metric, indicating the extent of the dispersion among machines. 

\noindent
\textbf{Step 2: Prioritization of the monitoring metrics.} Based on the maximum Z-score for each monitoring metric, \name uses a decision tree \cite{meng2020interpreting,guidotti2018survey} to prioritize the sensitivity of each metric in identifying the faulty machine. We resort to a decision tree for two primary reasons. Firstly, the logical structure of decision trees bears a resemblance to rule-based policies that are common in networking monitoring systems. For example, certain monitors utilize simple threshold rules, such as when \texttt{CPU Usage} drops to nearly zero \cite{zhao2017advanced}. Secondly, decision trees offer high expressiveness and faithfulness, attributed to their lack of parameters and the ability to represent complex decision-making \cite{blockeel1998top}.

To construct a decision tree, \name gathers the maximum Z-score for each metric from step 1 as an individual instance for the time window of the training task. The instance is labeled manually as normal or abnormal depending on whether a faulty machine exists within this window. Instances across multiple time windows and multiple training tasks are used together to train a decision tree.

\begin{figure}[tb]
    \vspace{0.0in}
	\centering
	\includegraphics[width=0.85\columnwidth]{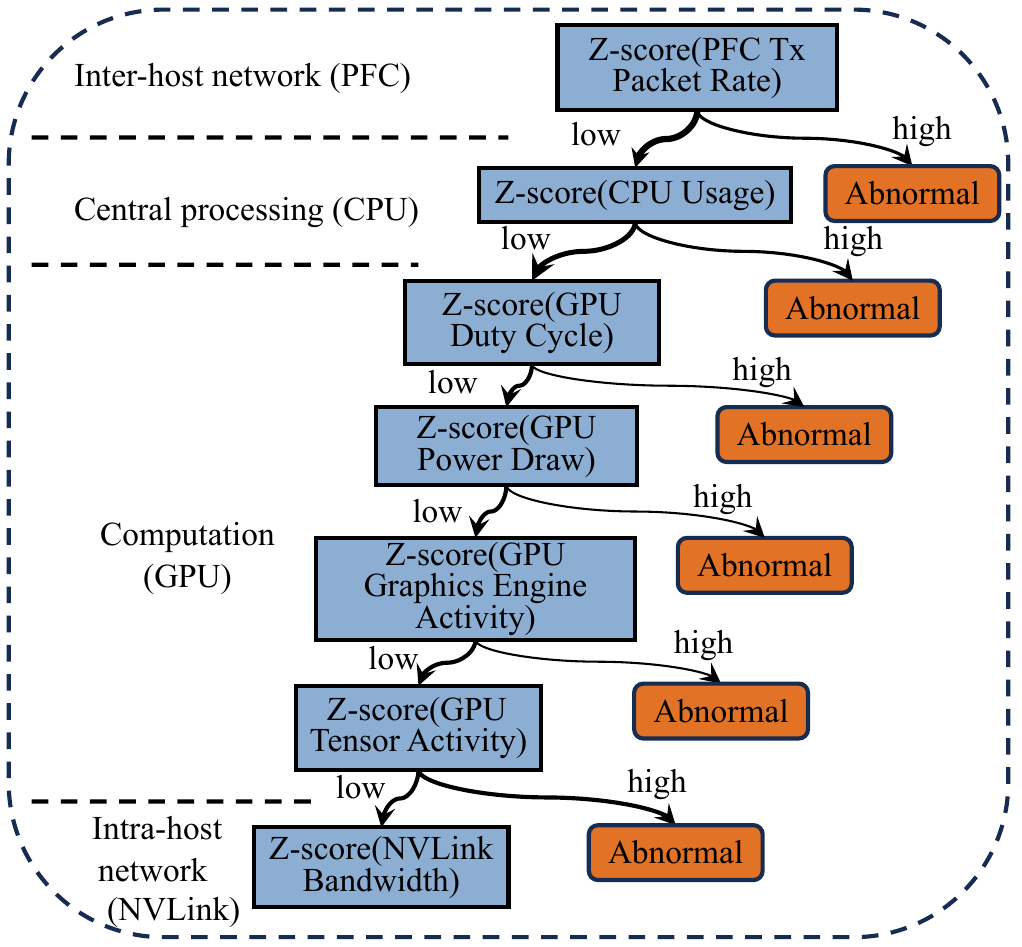}
	\vspace{-0.1in}
	\caption[12]{Top 7 layers of the decision tree for prioritization.}
	\vspace{-0.2in}
	\label{fig:design-tree}
\end{figure}

As shown in Figure~\ref{fig:design-tree}, monitoring metrics are prioritized based on their sensitivity to faults. The decision tree employs a step-by-step approach to classify the instances by analyzing the Z-scores of each metric. Nodes located closer to the root of the tree indicate that the corresponding monitoring metrics are more sensitive to the occurrence of a faulty machine. PFC, CPU, GPU, and NVLink-related metrics are identified as the most informative ones. Specifically, \texttt{CPU Usage} is relevant to running process states, while the four GPU metrics relate to the computation states. Additionally, \texttt{NVLink Bandwidth} and \texttt{PFC Packet Rates} are indicative of intra-host and inter-host network quality. The outcome aligns with Table~\ref{failure_type}, where CPU and GPU enjoy the highest priority.

\subsection{Online Faulty Machine Detection}
\label{subsec:detection}
\noindent
\yangtao{During run-time detection, \name leverages the concepts of machine-level similarity (\ref{subsec:similarity}) and continuity (\ref{subsec:continuity}), since the faulty machine tends to exhibit an abnormal pattern over a period. Given the data from each machine, \name follows the order of the decision tree prioritization results and selects a metric. The metric data is then denoised by the correlated model. \name then runs step 1\&2 introduced below for detection. If none is detected, \name picks the next metric and repeats step 1\&2 until one is detected. Fortunately, if no one is detected after passing all the models, \name assumes no anomaly occurs up to this time.}

\noindent
\textbf{Step 1: Similarity-based distance check per time window.} To detect a faulty machine during a time window, \name compares the similarity (\ref{subsec:similarity}) among all the machines for the same metric. In the presence of a faulty machine, its denoised data from LSTM-VAE is distinguishable from others.

To initiate the detection for a time window, the monitoring data from each machine (\eg a $1 \times w$ vector for a machine) is fed into the corresponding model successively. The reconstructed embedding for the $i$-th machine is captured by \name for the following distance calculation. Specifically, 
\name calculates the pairwise Euclidean distances of embeddings between every two machines, as it expresses distinct differences between normal and abnormal samples and provides characterizations of various record types \cite{weller2014survey}. For each machine, \name calculates the sum of the distances to other machines, representing its dissimilarity. Since the distance magnitude shifts with machine scales, we calculate the normal score for each sum value of the machines to normalize. The machine with the maximum normal score is probably the faulty one. If the maximum normal score is higher than a \textit{similarity threshold}, the machine is assumed as a candidate of the time window.

\noindent
\textbf{Step 2: Continuity check for consecutive time windows.} The detected candidate of a time window might be a false alarm due to instant bursts or temporary counter noises, so the idea of continuity (\ref{subsec:continuity}) is essential. This is because faults often lead to deteriorated performance for a period. \name shifts the time window with a stride of one to detect the potentially faulty machine for new windows. If the same machine is detected with consecutive times that exceed a \textit{continuity threshold}, it is considered a truly faulty machine. A proper \textit{continuity threshold} can be set as 4 minutes as it is adequate to filter out short-term noises and will not exceed the typical lasting time of a deteriorated performance in Figure~\ref{fig:motivation-pfc}.
\vspace{-0.05in}
\section{Implementation}
\label{sec:implementation}
\vspace{-0.05in}

\noindent
\yangtao{\name has been deployed in our ML system for a year. It runs on a dedicated machine with two dual-port ConnextX-6 25G-RNICs~\cite{connectx-6-dx}, 128 Intel Xeon Platinum 8336C CPUs, 512G memory, and 1.6T disk. The high-speed RNICs ensure the fast transmission of monitoring data, while computation and storage resources are adequate for real-time detection.}

\yangtao{\name monitors all the ongoing training tasks throughout their life cycles in our production environment. For a task, \name is called at pre-determined intervals (\eg every 8 minutes). Upon a call, \name pulls 15-minute data for the metrics listed in Appendix~\ref{sec:Appendix} from a database for all machines associated with the task. The metrics cover aspects of computation, communication, and storage. The database updates monitoring data per second from all the machines. If \name identifies a faulty machine, an alert is triggered to a driver and relevant engineers. After the driver submits the machine IP to be blocked and the Pod information to Kubernetes, the faulty machine will be evicted and replaced by a new one, before a fast recovery from recent checkpoints \cite{jiang2024megascale}. Importantly, the running of \name will not interfere with online distributed training tasks, as \name works as a backend service.}

\noindent
\yangtao{\textbf{Task workload.} The monitored training tasks are distributed across four to over 1000 machines, with GPU numbers up to more than 10000 for a task. Concurrent tasks could be monitored by \name. Each task is running on machines of homogeneous GPU and RNIC architectures on rail-optimized topology with up to three layers of switches. TP for computation, PP for gradient calculation and propagation, and DP that performs gradient synchronization are efficiently used for our LLM pre-training. These 3D parallelism strategies \cite{jiang2024megascale,shoeybi2019megatron} (\ref{subsec:similarity}) facilitate balanced computation (GPU usage, power, temperature \etc), storage (memory usage \etc), and communication (intra-host and inter-host throughput) across machines. Models trained in our ML system include the Transformer \cite{vaswani2017attention} and so on. The model sizes range from under 32B to over 500B.}
\vspace{-0.1in}
\section{Evaluation}
\label{sec:evaluation}
\vspace{-0.1in}

\noindent
\yangtao{\textbf{Dataset.} Our dataset includes 150 run-time fault instances from online distributed training tasks with 3D parallelism \cite{jiang2024megascale}. The dataset includes instances over nine months. Each task involves 4 to over 1500 machines (up to 10,000 NVIDIA Ampere GPUs), covering all the scale groups in Figure~\ref{fig:motivation-machine_scale_failure_times}. 30\% of the tasks involve a minimum of 600 machines. All fault types listed in Table~\ref{failure_type} are covered. The dominant ones are ECC error (25.7\%), CUDA execution error (15\%), GPU execution error (10\%), and PCIe downgrading (8.6\%). Our dataset focuses on faults in an individual machine, as they account for 99\% of all faulty cases in the production environment (\ref{subsec:multiple_faulty_machines} for concurrent faulty machine evaluation). The monitoring metrics in Appendix~\ref{sec:Appendix} were collected at the second-level granularity. Due to the fast eviction and recovery process, verifying whether all evicted machines are faulty is challenging. Consequently, our dataset encompasses the run-time instances where the actual faulty machine could be manually confirmed via offline log-checking, nccl-tests, or hardware tests (\eg~\cite{liu2023hostping}). For LSTM-VAE training, we use data from the first three months and the rest for evaluation.}

\noindent
\textbf{Metrics.}
For a task, we denote true positives (TP) as the correct machine detection following a fault, and false negatives (FN) as errors in machine detection or missed detections during a fault. True negatives (TN) refer to the correct approvals when machines are running normally, while false positives (FP) refer to false detections when there is no fault. Then we calculate Precision, Recall, and F1-score as our metrics.

\begin{figure}[tb]
	\centering
	\includegraphics[width=0.89\columnwidth]{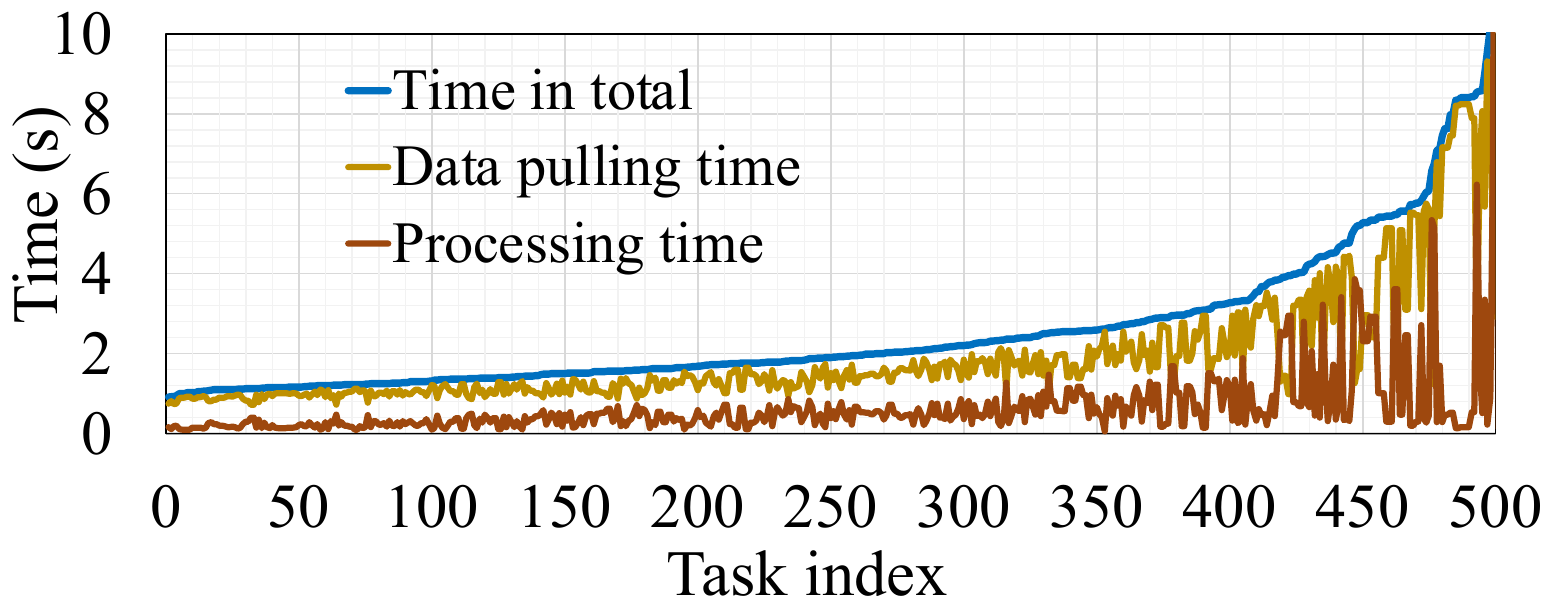}
	\vspace{-0.1in}
	\caption[12]{The total data processing time for a call of \name.}
	\vspace{-0.2in}
	\label{fig:evaluation-time}
\end{figure}

\begin{figure}[tb]
	\vspace{0.05in}
	\centering
	\includegraphics[width=0.88\columnwidth]{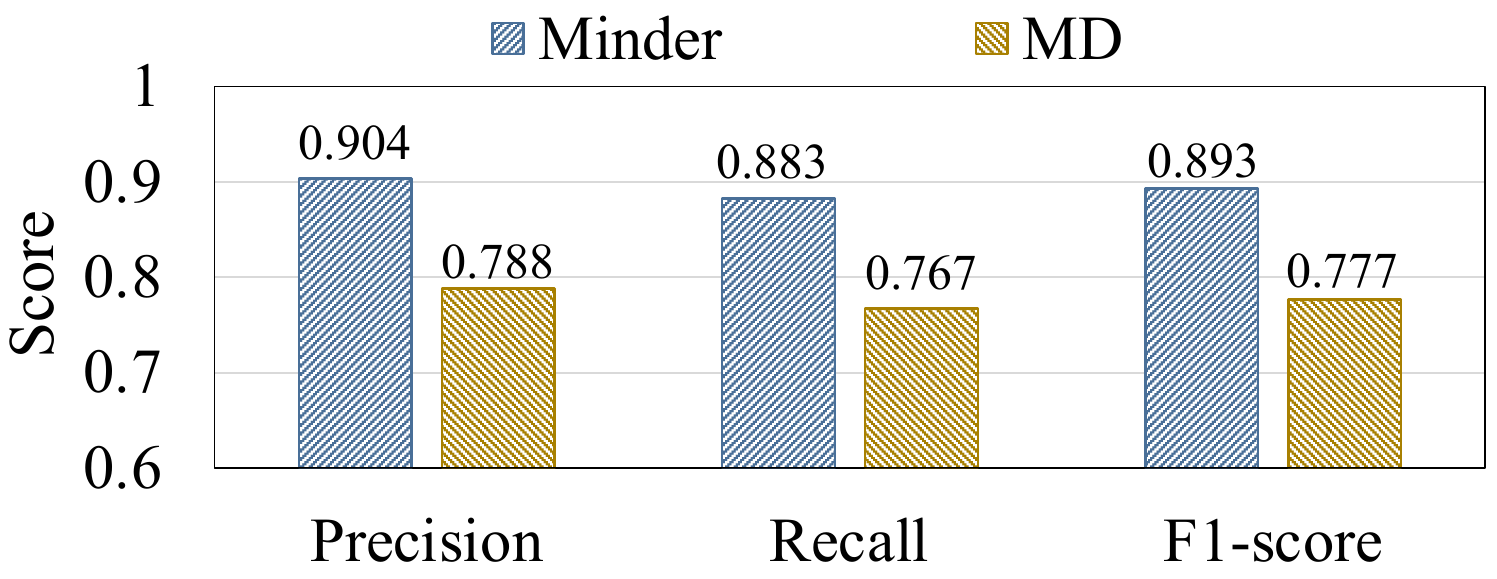}
	\vspace{-0.1in}
	\caption[12]{Comparison with a baseline algorithm MD \cite{ghorbani2019mahalanobis, leys2018detecting}.}
	\vspace{-0.2in}
	\label{fig:evaluation-overall}
\end{figure}

\begin{figure*}[tb]
	\centering
	\includegraphics[width=1.85\columnwidth]{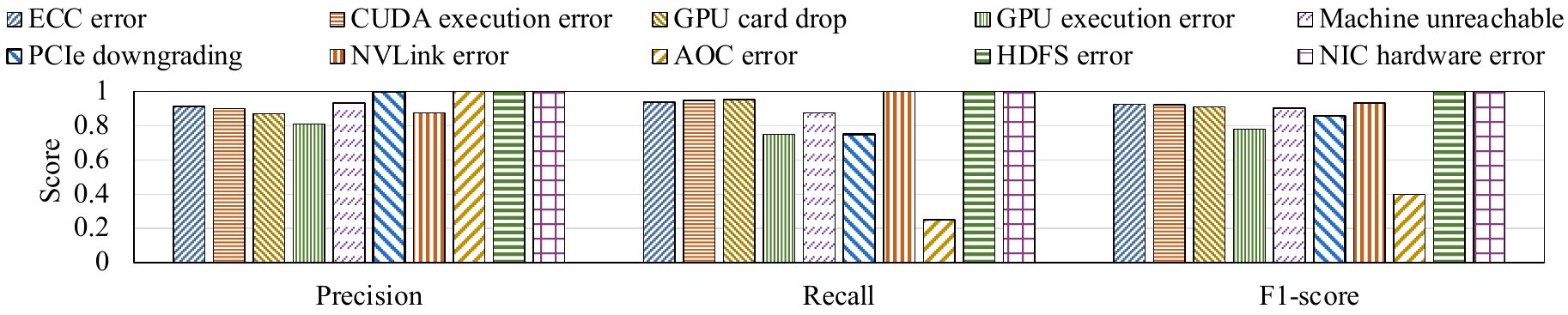}
         \vspace{-0.1in}
	\caption[12]{\yangtao{Accuracy for various fault types.}}
	\vspace{-0.2in}
	\label{fig:evaluation-types}
\end{figure*}

\subsection{Overall Performance}
\label{subsec:overall_detection_performance}
\noindent
\textbf{Total data processing time.} In Figure~\ref{fig:evaluation-time}, a call of \name takes 3.6 seconds on average to make an alert. It includes data pulling time (fetching the pertinent monitoring data from Data APIs) and processing time (preprocessing, and running inference for faulty machine detection). Due to \name's deployment on an exclusive machine, it reduces the time by 99\% (shorter time by 500 $\times$ compared with Figure~\ref{fig:motivation-manual_time}) if engineers otherwise manually inspect machine information one by one.

\noindent
\textbf{Comparison with the baseline.} \yangtao{The baseline for comparison is Mahalanobis Distance (MD) \cite{ghorbani2019mahalanobis, leys2018detecting, mahalanobis2018generalized}. MD is widely used in identifying outliers.} It considers the variable correlations in multi-dimensional data and calculates features like mean, variance, skewness, and kurtosis before applying principle component analysis (PCA) and computing the pairwise distances. We keep other processes the same for comparison.

In Figure~\ref{fig:evaluation-overall}, the precisions are 0.904\&0.788, with recalls of 0.883\&0.767, leading to F-1 scores of 0.893\&0.777 for \name and MD. These findings demonstrate that \name effectively detects actual faults and lowers the rate of false alarms. MD detects the outlier machine from the statistical perspective but exhibits lower accuracy, indicating that jitters and noises interfere with statistical features. \name outperforms MD by using LSTM-VAE for denoising and extracting the data patterns for a better distance calculation.

\noindent
\yangtao{\textbf{Performance breakdown with fault types.} By considering individual fault types in Table~\ref{failure_type}, the results slightly differ. In Figure~\ref{fig:evaluation-types}, \name effectively handles faults like ECC error, CUDA execution error, GPU card drop, machine unreachable, NVLink error, HDFS error, and NIC hardware error. These faults are relative to CPU, GPU states, and networking performances that \name monitors.}

\yangtao{GPU execution error and PCIe downgrading present a lower recall, since some faulty instances have concurrent faulty GPUs or PCIe links within a machine. Owing to the 3D parallel topology, faults swiftly impact multiple machines in both the DP and PP groups, leading to an instant group effect. Thus, the time-series granularity at the second level is insufficient for \name. Other fault types, such as AOC errors, are partially missed due to the current absence of optical cable-related counters for capturing useful monitoring metrics. We will continue to have more hardware counters in the future. Overall, AOC errors occupy only a small portion, so the overall accuracy is still high.}

Upon examining the errors made by \name, most of them were not entirely incorrect. In many error cases, the machine incorrectly detected showed short-term metric fluctuations or performance jitters before being normalized afterward. These detected fluctuations, however, were not the root cause of the task halt. Nonetheless, these errors should not be ignored as they led to short-term jitters in the training performance.

\begin{figure}[tb]
	\centering
	\includegraphics[width=0.9\columnwidth]{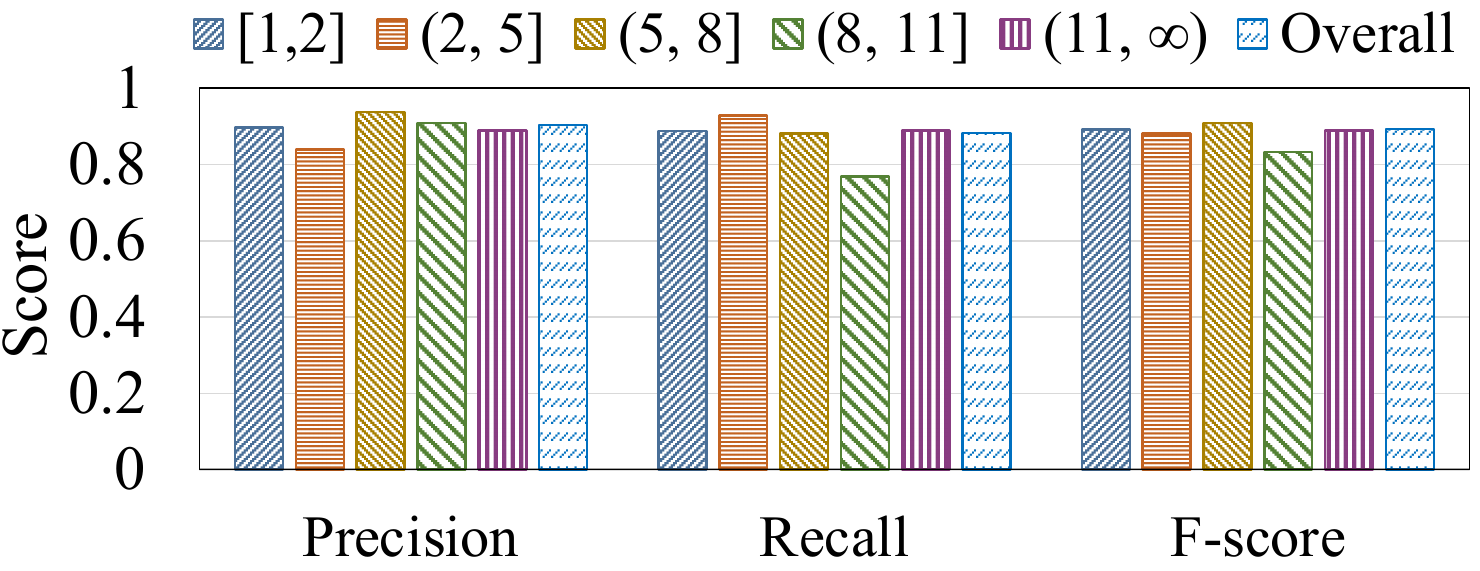}
	\vspace{-0.05in}
	\caption[12]{\yangtao{Accuracy for tasks with varied lifecycle fault occurrences.}}
	\label{fig:evaluation-multi-failure}
   \vspace{-0.2in}
\end{figure}

\noindent
\yangtao{\textbf{Performance breakdown by fault occurrences of a task's lifetime.} Training tasks with various workloads experience distinct occurrences of faults. Figure~\ref{fig:evaluation-multi-failure} categorizes the performance relative to the fault numbers throughout a task’s lifetime, which may span up to months for extensive workloads. In our dataset, 70\% of the tasks display no more than five faults, whereas over 15\% face more than eight faults throughout their lifetime. Given the lack of interdependence among individual faults, the accuracy is not tied to the fault occurrences. Since faults are random and a faulty machine will be promptly auto-replaced in the production environment, the fault occurrences do not affect performance. Despite the lower recall for the (8, 11] group, it primarily originates from the limited task number of this group.}

\subsection{Analysis of Monitoring Metric Selection}
\label{subsec:analysis_of_kpi_selection}
\noindent
To validate the proper selection of monitoring metrics in (\ref{subsec:kpi_selection}), we conduct experiments to show that fewer or more metrics do not improve accuracy. Notably, most of the remaining metrics not used by \name are GPU-related, including \texttt{GPU Temperature}, \texttt{GPU Clocks}, \texttt{GPU Memory Bandwidth Usage}, and \texttt{GPU FP Engine Activity}. Thus, we use only \texttt{GPU Duty Cycle} to train a GPU model with fewer metrics, while adding these unused GPU-related monitoring metrics to train a GPU model with more metrics. We keep the other settings unchanged for the comparison.

\begin{figure}[tb]
	\centering
	\includegraphics[width=0.85\columnwidth]{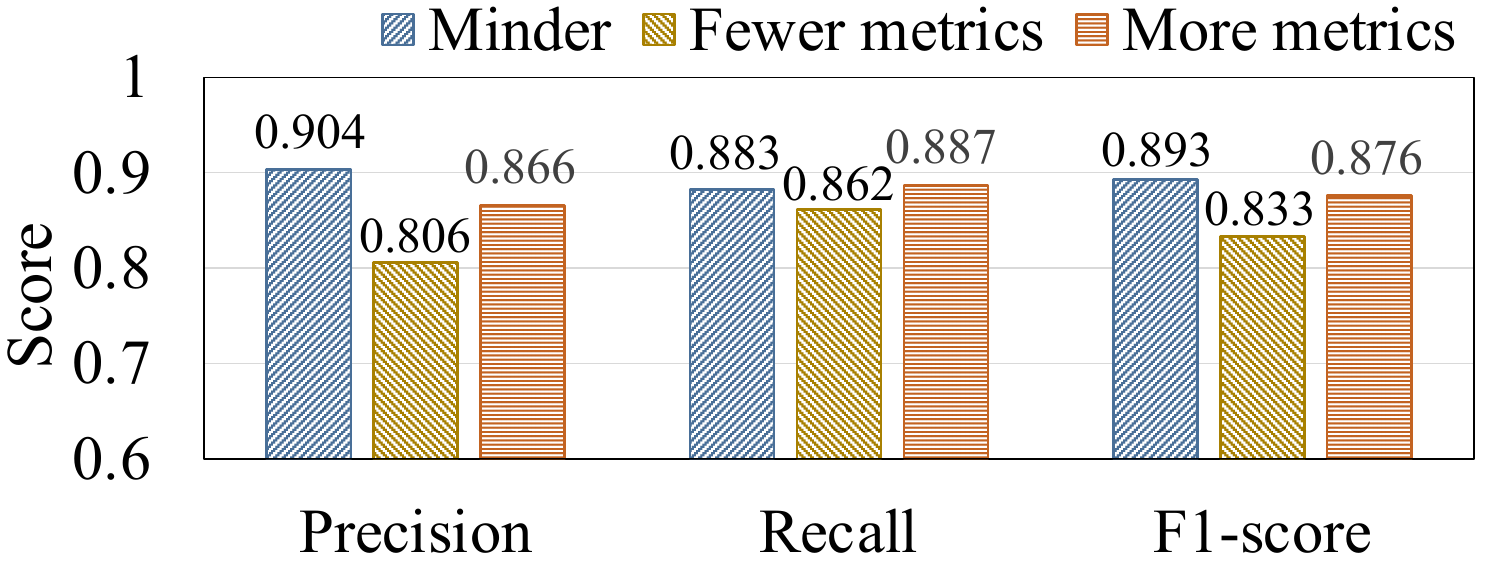}
	\vspace{-0.05in}
	\caption[12]{Comparison with different metric selections.}
	\vspace{-0.16in}
	\label{fig:evaluation-metric}
\end{figure}

Figure~\ref{fig:evaluation-metric} reveals that including more monitoring metrics achieves a higher recall, but its precision is lower. More metrics may introduce mutual interference, since different metrics may indicate different patterns that will obfuscate the detection. On the other hand, using fewer metrics undermines outlier detection capacity due to the exclusion of key metrics. Our optimal selection of metrics achieves the best precision, meaning that \name alerts far fewer false alarms than others. Based on the metric priority result (\ref{subsec:kpi_selection}), the top metrics are sufficient to cover all components that might malfunction.

\subsection{Analysis of Model Selection}
\label{subsec:model_comparison}
\noindent
We contrast \name against other statistical methods or model variations to show the proper choice of LSTM-VAE. A simple approach is calculating the Euclidean Distances of the preprocessed raw data (RAW) without using VAE. Variants of LSTM-VAE include concatenating the embeddings of all the models as a whole for distance calculation (CON) or training an integrated LSTM-VAE model with all the monitoring metrics (INT).

\begin{figure}[tb]
	\vspace{0.0in}
	\centering
	\includegraphics[width=0.85\columnwidth]{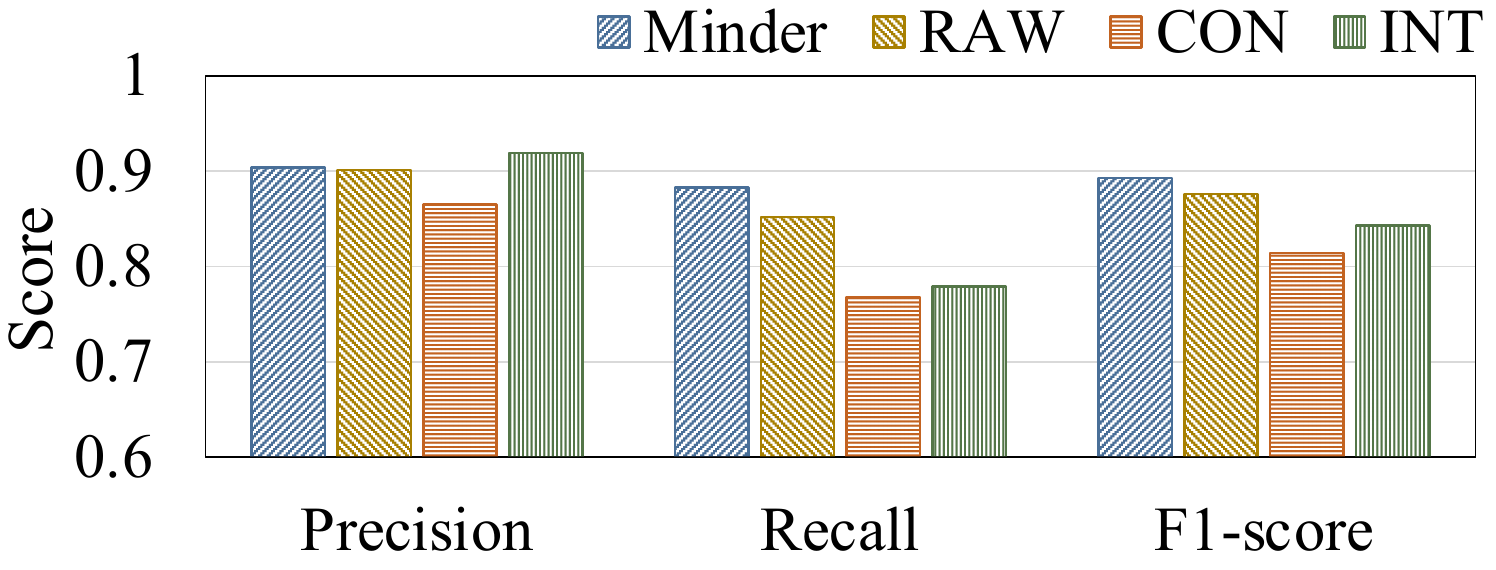}
	\vspace{-0.05in}
	\caption[12]{Comparison with different model selections.}
	\vspace{-0.2in}
	\label{fig:evaluation-model}
\end{figure}

In Figure~\ref{fig:evaluation-model}, \name outperforms others in recall and F1-score. Their recalls vary. Worse recall generated by RAW illustrates the significance of eliminating noises. \name, in contrast, reconstructs for denoising. CON and INT show worse recall because they consider multiple metrics with equal significance by calculating distances from evenly concatenated embeddings or regarding all the metrics as a whole for input. However, not all monitoring metrics have an equal sensitivity to faults. Mutual interference exacerbates the performance of CON and INT. \name, on the other hand, leverages the VAE for denoising and separates metrics to enhance the performance. \yangtao{Besides, comparing the input and reconstructed data of LSTM-VAE yields a Mean Squared Error (MSE) lower than 0.0001, demonstrating effective reconstruction.}

\subsection{Analysis of Continuity and Threshold}
\label{subsec:analysis_of_continuity}
\begin{figure}[t]
	\centering
	\includegraphics[width=0.85\columnwidth]{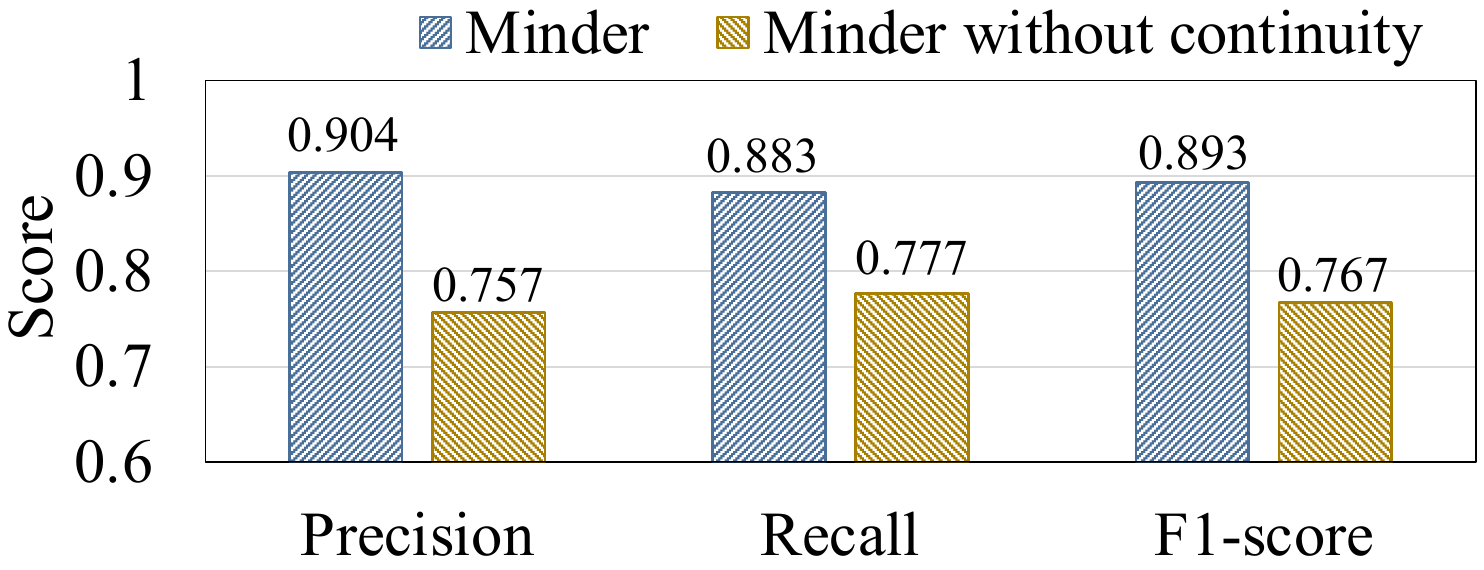}
	\vspace{-0.05in}
	\caption[12]{Accuracy with/without continuity.}
	\vspace{-0.2in}
	\label{fig:evaluation-continuity}
\end{figure}

\noindent
\yangtao{To verify the feasibility of continuous detection, we compare \name without the application of continuity (\ref{subsec:continuity}). \name ensures the same machine is detected multiple times. Without continuity, an alert will be made immediately upon a fault detection during the time window. The results in Figure~\ref{fig:evaluation-continuity} imply that the overall performance is worse without continuity. More false alarms are triggered due to occasional short-term jitters. The continuity emphasizes the degraded performance for a period caused by a fault's gradual impact on other machines. Thus, \name filters discrete bugs, noises, or jitters.}

\yangtao{Note that we set the \textit{continuity threshold} (\ref{subsec:detection}) as four minutes to reduce false alarms caused by jitters or noises. That means \name only alerts when a detected machine endures dissimilarity for four minutes. The threshold is chosen empirically based on the fault duration in Figure~\ref{fig:motivation-duration}. Most of the faults last longer than four minutes before the halt. A shorter continuity threshold introduces more false alarms, while a longer one excludes more actual faulty detection results.}

\subsection{Choice of Distance Measures}
\label{subsec:choice_of_distance_measures}
\noindent
We compare \name’s pair-wise Euclidean Distance algorithm with Manhattan Distance (MhtD) and Chebyshev Distance (ChD). MhtD adds up the absolute distances from each dimension of the embeddings while ChD uses the largest difference in their coordinates. We replace the distance algorithm to rerun the experiments. 

\begin{figure}[t]
	\centering
	\includegraphics[width=0.85\columnwidth]{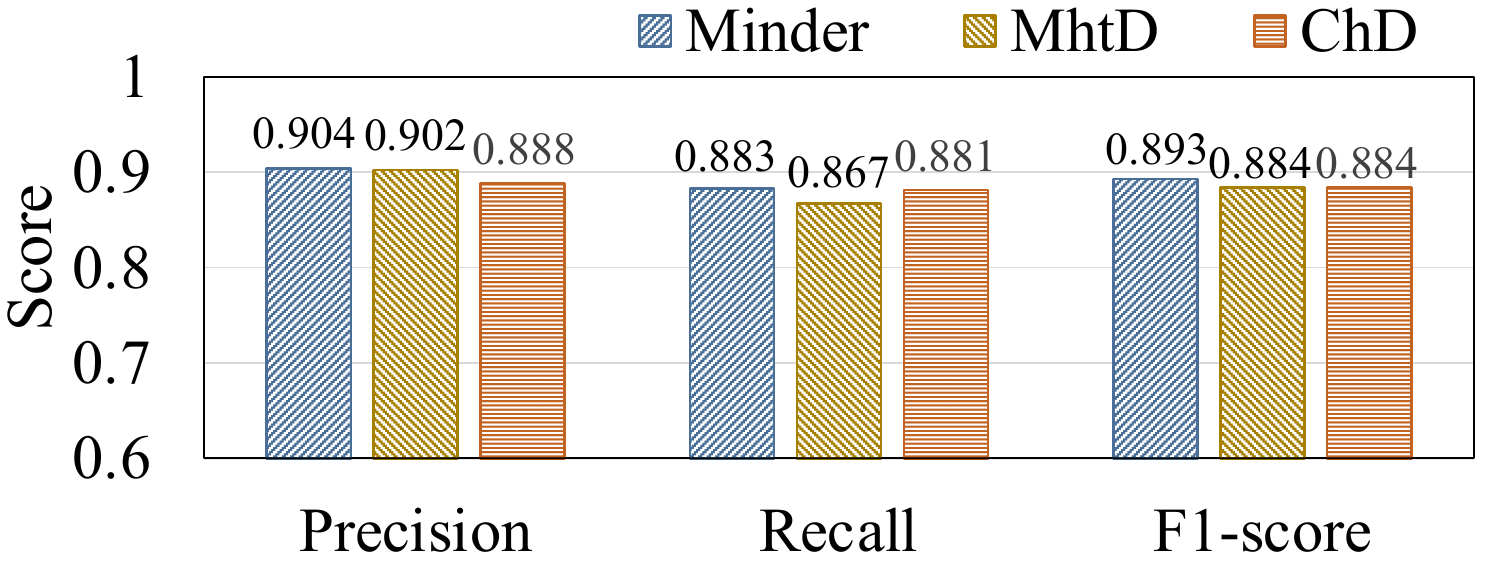}
	\vspace{-0.05in}
	\caption[12]{Comparison with different distance measures.}
	\vspace{-0.2in}
	\label{fig:evaluation-distance}
\end{figure}

In Figure~\ref{fig:evaluation-distance}, \name achieves similar performance to others, suggesting that the embeddings from LSTM-VAE are already representative and the outlier is distinctive for any distance calculation method. The comparison with MhtD implies that spatial distribution is a solid representation because both methods use the distances of multiple dimensions. ChD's worse precision suggests that a single spatial difference is insufficient for comparing dissimilarity. \name considers the overall distance from the outlier to other normal ones and the dissimilarity will be intensified by summing the distances.

\subsection{Performance with Multiple Concurrent Faulty Machines}
\label{subsec:multiple_faulty_machines}
\noindent
\yangtao{Analyzing \name's ability to handle multiple faults is intuitive. Such detection capability largely depends on the faulty machine scale ratio and the granularity of monitoring data. Given the 3D parallelism \cite{jiang2024megascale}, a machine participates in multiple DP and PP groups. More faulty machines impact more groups, inducing faster negative propagation across the entire cluster. The data granularity determines the visibility of the propagation process. With a brief duration of less than 100 milliseconds per interaction, the dissimilar pattern may be overlooked due to coarse-grained monitoring. Yet, millisecond-level monitoring is not widely deployed due to the high overhead.}

\yangtao{In our production environment, concurrent faulty machine instances only occur due to automatic switch reboots or switch-related AOC errors. A switch reboots in response to high temperatures, extreme port congestion, and so on. Thirty-two connected machines will be forced to go offline out of a total of 600 machines in our environment. However, \name hardly distinguishes the faulty outliers. Firstly, the fault ratio is relatively high. Given our rail-optimized topology and 3D parallelism mechanism, communication among 32 machines contains at most 256 DP groups, quickly propagating across other machines. Moreover, the current second-level time granularity monitoring limits \name's ability. The rapid spread is only observable at the millisecond level following several training interactions. However, the rate of such multiple faulty machine instances is less than 1\%, sometimes zero a month. Should a switch reboot, the switch monitoring system will automatically alert the engineers.}

\begin{figure}[tb]
	\centering
	\includegraphics[width=0.9 \columnwidth]{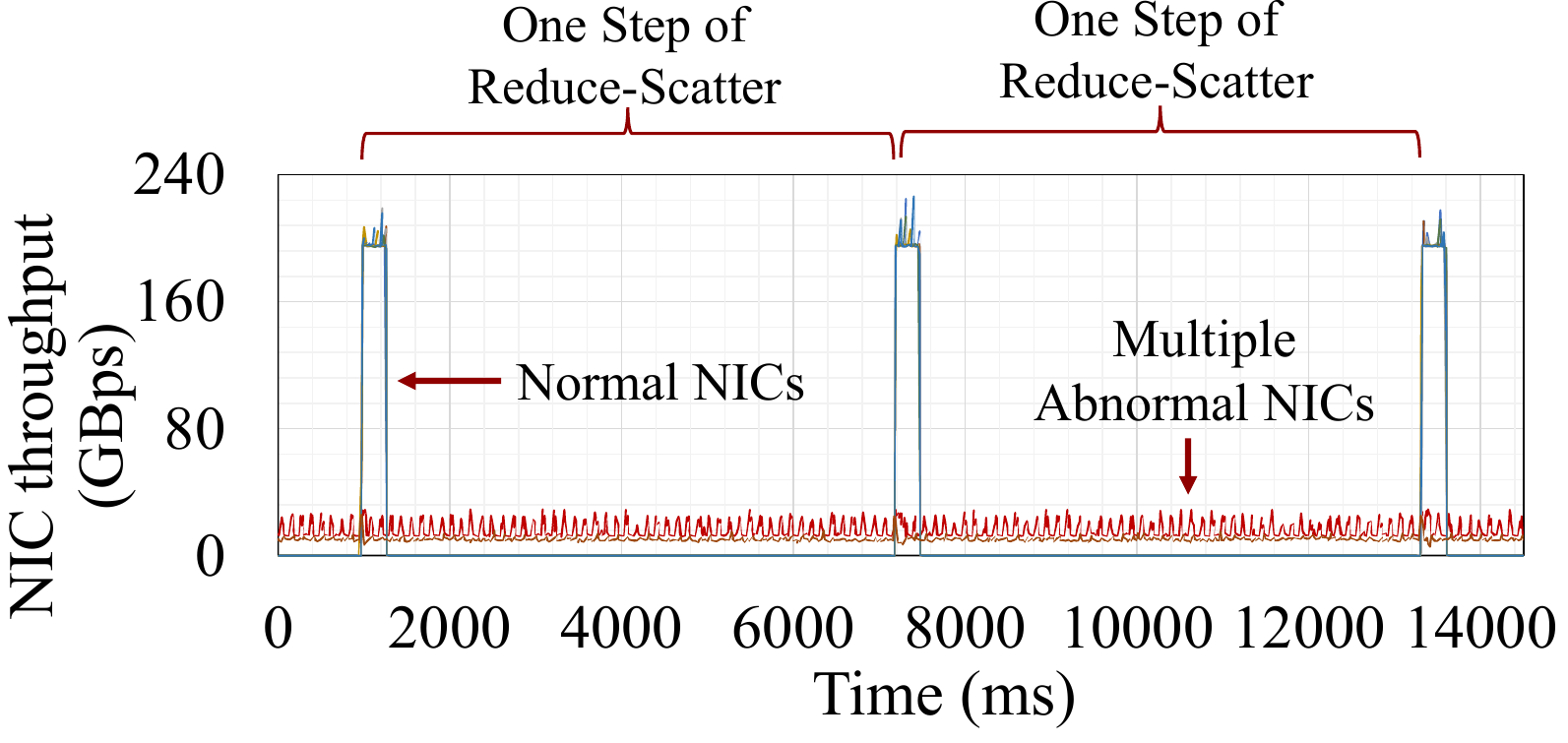}
	\vspace{-0.1in}
	\caption[12]{\yangtao{Millisecond-level NIC throughput for all machines after injection of PCIe downgrading on two NICs.}}
     \vspace{-0.2in}
	\label{fig:evaluation-concurrent_faulty_machines}
   \vspace{0.0in}
\end{figure}

\yangtao{To demonstrate \name's ability to detect multiple faulty machines, we injected PCIe downgrading into two of four machines simultaneously, with customized millisecond-level monitoring. Thus, the ratio of faulty machines is higher and the data granularity is finer than the switch-reboot instance. In the experiment, each machine was equipped with eight NVIDIA Ampere GPUs, running Reduce-Scatter collectively. Two PCIe links on two machines were purposely degraded. With the millisecond-level data from the NICs, \name could detect the two NICs connected to the faulty PCIe links. These two NICs presented the largest outlier distances during Reduce-Scatter. Figure~\ref{fig:evaluation-concurrent_faulty_machines} shows the millisecond-level monitoring, where normal NICs demonstrate a high throughput at the beginning of each Reduce-Scatter step to transmit their data to the next node. In contrast, the NICs with downgraded PCIe links exhibit steady and low throughput. Thereafter, normal ones drop to zero, after transmitting their data, waiting for the slow NICs for synchronization. \name can capture this dissimilarity, thanks to the granularity of milliseconds. Our injection experiments also show \name's ability to detect other concurrent faults, such as GPU degradation and NIC throughput downgrading.}

\vspace{-0.15in}
\section{Discussion}
\label{sec:discussion}
\vspace{-0.07in}

\noindent
In this section, we discuss the lessons we learned during implementation and potential future works.

\noindent
\yangtao{\textbf{\name and other monitoring tools.} Large-scale model training involves the cooperative efforts of multiple teams. Other monitoring tools used along with \name include switch state monitoring, periodic heartbeat messages (IP, hardware states, Pod names \etc), RDMA traffic down-limit alerts, R-Pingmesh \cite{R-Pingmesh} (a pingmesh-like \cite{guo2015pingmesh} connection testing), and automatic text analysis for GPU error detection. As described in \ref{sec:implementation}, a detected machine will be replaced before restarting the task. These approaches primarily target network connection and jitters or GPU states, whereas \name provides metrics covering computation, storage, and communication resources. The combined use of them enhances the detection efficacy. Meanwhile, offline testing tools, such as DCGM \cite{dcgm}, EUD \cite{eud}, are used for intra-host bottlenecks diagnosis, though not feasible for run-time fault identification.}

\new{Similar systems and tools are introduced. For example, SuperBench \cite{298589} is a proactive system to ensure cloud AI infrastructure reliability. It runs model training and component benchmarks to identify incremental performance degradation on defective machines. The first difference is that it works proactively  because incremental performance degradation exists due to hardware redundancy. However, such gradual degradation process before a fault occurs is hardly observed in our distributed training. Second, SuperBench mainly improves the hardware-side reliability. Real-time software errors also occur frequently during model training. Besides, SuperBench  performs offline validation by running benchmarks. \name otherwise monitors the tasks throughout their life cycles. Thus, the integration of such a proactive system with \name is a more promising debugging solution. By executing benchmarks on the machines in a job and monitoring their running status, faults might be largely prevented and rapidly detected.}

\noindent
\yangtao{\textbf{\name's Generality.} \name can be extended in data granularity and the spectrum of available metrics. Second-level monitoring is currently deployed and used by \name. \name has demonstrated improved detection capability for concurrent faulty machines with millisecond-level monitoring data in \ref{subsec:multiple_faulty_machines}. With finer-grained data, the annihilated rapid propagation from the straggler will be revealed, as a training iteration only lasts tens of milliseconds. Besides, the currently available metrics for \name are out-of-band hardware counters. Other metrics (\eg AOC counters) and in-band traces (Torch Profiler \cite{torch_profiler}, the Megatron-LM timer, or CUDA event timer \cite{jiang2024megascale}) could also be utilized by \name. These new traces offer fine-grained training and collective communication operation information for comprehensive performance monitoring.}

\noindent
\textbf{\name's robustness of other faults.} Notably, \name is allowed to detect new or rare fault types not covered in training, as long as the monitoring data presents discernible dissimilarities. For concurrent faulty machines, provided multiple machines present a distinctive dissimilarity, \name can detect them simultaneously.

\noindent
\textbf{Machine-level similarity.} 
\yangtao{Large models such as those in LLM training and multi-modal tasks often employ 3D parallelism. This results in a trend of consistency across machines for computation, communication, and storage, allowing \name to detect faulty machines. As a result, we focus on machine-level detection instead of finer granularity. 
We also consider \name in other workloads, such as large-scale inference and fine-tuning. As long as these workloads satisfy the requirements of inter-machine metric similarity and fault continuity, \name could be applied. Future work will explore \name's effectiveness in other workloads.}

\noindent
\textbf{Not all failed tasks have the right label.} 
Although the labeled machine is the root cause, the \name-detected machine may also have temporary performance fluctuations.
It is necessary to inform engineers of such performance jitters.

\noindent
\textbf{Root cause analysis.} \name detects faults at the machine level. The root cause for a fault indicated by a metric is uncertain. Extra labor is employed for further network jitter and short-term straggler analysis. In the future, we plan to design fine-grained run-time monitoring for root cause identification.

\vspace{-0.15in}
\section{Related Work}
\label{sec:related_work}
\vspace{-0.07in}

\noindent
In this section, we briefly introduce related works on anomaly diagnosis\cite{arzani2017closing, ghita2013toward, heller2013leveraging, herodotou2014scalable,katz2010reverse, matteson2014nonparametric, mysore2014gestalt, tammana2016simplifying, zhu2015packet, gong2020microscope, geng2019simon}.

\noindent
\textbf{Intra-host diagnosis.} 
Run-time diagnosis detects anomalies without disturbing the running tasks. Leveraging the existing counters\cite{Nvidia_counter,Intel_counter} from the machine is a direct method\cite{sun2021ctf,xu2018unsupervised,su2019robust}. For example, BRCA\cite{nie2016mining}, MonitorRank\cite{kim2013root}, and FChain \cite{nguyen2013fchain} construct a dependency graph based on historical monitoring data or traces for anomaly alerts. These DL algorithms have been proven to be automatic, robust, and flexible. Another naive but effective way is to check logs or \texttt{dmesg}\cite{dmesg} using natural language processing (NLP)-based methods \cite{anvik2006should}. However, log-based approaches are limited to their log content and information processing abilities.

Offline diagnosis requires specific tools to detect possible intra-host bottlenecks when the machine is not running tasks. Liu \etal \cite{liu2023hostping} and Martinasso \etal \cite{martinasso2016pcie} implement tools to test if there is NVLink\cite{NVLink} or PCIe link\cite{PCIe} degradation or congestion. Collie \cite{kong2022collie} is a "fuzzing"-like implementation to help uncover potential performance bottlenecks in RNICs. Deepview \cite{zhang2018deepview} is designed for virtual hard disk failure localization. These approaches are useful ahead of running tasks, so they cannot be directly used during the large-scale training process.

\noindent
\textbf{Inter-host diagnosis.} 
NetBouncer\cite{tan2019netbouncer} leverages the IP-in-IP technique to actively localize failure devices or links among millions of servers in a data center network. Similarly, SNAP \cite{yu2011profiling} monitors TCP statistics and socket logs for network diagnosis. Pingmesh \cite{guo2015pingmesh}, Haecki\cite{haecki2022diagnose}, and Fathom\cite{qureshi2023fathom} monitor end-to-end latencies between arbitrary servers or detailed RPC performance in data centers. Nonetheless, they only detect failures along the routes instead of intra-host hardware failures that degrade the training speed. Cloud system diagnosis\cite{li2023conan,zeng2023traceark, harsh2023murphy, gan2021sage} are based on contextual data patterns and associations of microservices on multiple machines. Such patterns and dependencies are eliminated in distributed training, where machines exhibit similar workloads.

\noindent
\textbf{Algorithms for anomaly detection and diagnosis.} 
The first type is statistics-based methods. Apart from Euclidean distance, Pearson Correlation \cite{cohen2009pearson}, Kendall’s tau\cite{kendall1938new}, and Spearman Correlation\cite{zar2005spearman} also quantify the similarity between two vectors and discover the deviating anomalies. Setting parameters as thresholds from experienced operators is usually required \cite{liu2019fluxrank}.

Supervised algorithms are widely used for anomaly detection\cite{liu2015opprentice, laptev2015generic}. EGADS in Yahoo\cite{laptev2015generic} leverages diverse machine learning methods for large-scale univariate time series anomaly detection. Machine learning algorithms like random forest \cite{palczewska2013interpreting} are used for incident routing \cite{gao2020scouts}, VM compromise detection\cite{arzani2020privateeye}. However, supervised learning does not fit in our context, where the goal is to detect the faulty machine instead of classification.

Unsupervised learning \cite{liu2019fluxrank, li2018robust, li2019linear, su2019robust, sun2021ctf, xu2018unsupervised, 295713} identifies outliers from monitoring data for root cause detection and machine state detection, or enables
dialogue-based diagnosis chatting. Apart from VAE (\ref{subsec:training}), clustering \cite{liu2019fluxrank, li2018robust, li2019linear} is commonly used to cluster the machines with similar monitoring data change patterns or detect anomalies for time series.

%\vspace{-0.1in}
\section{Conclusion}
\label{sec:conclusion}
%\vspace{-0.08in}

\noindent
This paper presents \name, addressing the problem of faulty machine detection in distributed training tasks. \name leverages the concept of similarity among machines and the continuity of a fault during training. \name has been deployed in our production environment for a year to assist engineers in training diagnosis. Evaluation results demonstrate the reduced time required by \name and the effectiveness of its design choices for training tasks.

This work does not raise any ethical issues.

\vspace{-0.1in}
\section*{Acknowledgments}
\label{sec:acknowledgement}
\vspace{-0.05in}

\noindent
\new{We thank all the anonymous NSDI reviewers for their feedback that greatly improved the paper. We thank the broader ByteDance High-speed Network team and Applied Machine Learning team for their support throughout this project.}

\bibliographystyle{abbrv}
\bibliography{reference}

\clearpage
\appendix

\noindent
Appendices are supporting material that has not been peer-reviewed.
\vspace{-0.09in}

\section{Fault Types}
\label{sec:Appendix_introduction}
\noindent
The fault types are listed in Table~\ref{failure_type}.ECC error: caused by corrupted or lost data in (GPU) memory. PCIe downgrading: a link fault leading to a slow PCIe sending/receiving rate. NIC dropout: a NIC is missing from the OS. GPU Card drop: a disconnected GPU card.  NVLink error: a link fault between two Nvidia GPUs. AOC error: an error in high-speed active optical cables (AOC) on either the host network card or the switch side. CUDA execution error: an unexpected overflow or configuration leading to a failed CUDA program. GPU execution error: unexpected page-fault, out-of-memory, and other incorrect processing leading to GPU hang or other results. HDFS error: HDFS connection timeout, io error, and so on when loading or saving checkpoints. Machine unreachable: mostly due to malfunctioning SSH services or virtual machine services. Others: illegal memory access, failed scheduling, no disk storage, low resource usage, switch reboot, and so on.
    
\section{Collected Monitoring Metrics}
\label{sec:Appendix}
\noindent
Table~\ref{monitoring_metric} contains the monitoring metrics that we choose to collect in our production environment, though only a portion of them are used for training and detection. Other available host metrics could also be used by \name.

\begin{table}
  \small 
  \centering
  \caption{Monitoring metrics collected by \name.}
  \label{monitoring_metric}
  \vspace{-0.0in}
  \begin{tabular}{ll}
    \hline
    \textbf{Monitoring Metrics} & \textbf{Description} \\
    \hline
     \textbf{CPU Usage} & Percentage of CPU time being used.  \\
     \textbf{PFC Tx Packet Rate} &  Periodic counts of PFC packets sent by RDMA-enabled devices. \\
     \textbf{Memory Usage} & Percentage of memory being used.  \\
     \textbf{Disk Usage} &  Percentage of storage space being used on a disk. \\
     \textbf{TCP Throughput} & Periodic counts of the amount of TCP data being transmitted by a NIC.  \\
     \textbf{TCP+RDMA Throughput} & Periodic counts of the amount of TCP and RDMA data being transmitted by an NIC.  \\
     \textbf{GPU Memory Used \cite{GPU_memory}} & The amount of GPU memory being used by processes.  \\
     \textbf{GPU Duty Cycle \cite{GPU_memory}} &  Percentage of time over the past sample period when the accelerator is active.\\
     \textbf{GPU Power Draw} &  Periodic counts of the GPU power consumption. \\
     \textbf{GPU Temperature} &  The temperature of a GPU while it is operating, measured in degrees Celsius. \\
     \textbf{GPU SM Activity \cite{GPU_metrics}} & Averaged percentage of time when at least one warp is active on a multiprocessor. \\
     \textbf{GPU Clocks} & The clock speed of a GPU, reflecting the frequency of the GPU's processor. \\
     \textbf{GPU Tensor Core Activity \cite{GPU_metrics}} & Percentage of cycles when the tensor (HMMA / IMMA) pipe is active.  \\
     \textbf{GPU Graphics Engine Activity \cite{GPU_metrics}} & Percentage of time when any portion of the graphics or compute engines are active.   \\
     \textbf{GPU FP Engine Activity \cite{GPU_metrics}} & Percentage of cycles when the FP pipe is active.  \\
     \textbf{GPU Memory Bandwidth Utilization \cite{GPU_metrics}} &  Percentage of cycles when data is sent to or received from the device memory. \\
     \textbf{PCIe Bandwidth \cite{GPU_metrics}} & The rate of data transmitted/received over the PCIe bus.  \\
     \textbf{PCIe Usage \cite{GPU_metrics}} & Percentage of the bandwidth being used on the PCIe bus. \\
     \textbf{GPU NVLink Bandwidth \cite{GPU_metrics}} & The rate of data transmitted/received over an NVLink. \\
     \textbf{ECN Packet Rate} &  Periodic counts of ECN packets transmitted/received by a NIC. \\
     \textbf{CNP Packet Rate} &  Periodic counts of CNP packets transmitted/received by a NIC. \\
    \hline
    \vspace{-0.0in}
 \end{tabular}
\end{table}

\end{document}